\documentclass[aip,jcp, reprint]{revtex4-1} 
\usepackage{amsmath,amssymb,amsfonts, graphicx,color,multirow}

\newcommand{\hbe}{\mathbf{\hat{e}}}

\newcommand{\p}{\partial}
\newcommand{\rmd}{\mathrm{d}}

\newcommand{\bv}{\mathbf{v}}
\newcommand{\bJ}{\mathbf{J}}
\newcommand{\bj}{\mathbf{j}}

\newcommand{\tbv}{\tilde{\mathbf{v}}}
\newcommand{\tbJ}{\tilde{\mathbf{J}}}
\newcommand{\tbj}{\tilde{\mathbf{j}}}
\newcommand{\tbs}{\tilde{\mathbf{\sigma}}}
\newcommand{\tPsi}{\tilde{\Psi}}
\newcommand{\tc}{\tilde{c}}
\newcommand{\tr}{\tilde{r}}
\newcommand{\tX}{\tilde{X}}
\newcommand{\tU}{\tilde{U}}
\newcommand{\tE}{\tilde{E}}
\newcommand{\tmu}{\tilde{\mu}}
\newcommand{\tp}{\tilde{p}}
\newcommand{\tf}{\tilde{f}}
\newcommand{\talpha}{\tilde{\alpha}}
\newcommand{\hc}{\hat{c}}
\newcommand{\hU}{\hat{U}}

\newcommand{\tTheta}{\tilde{\Theta}}
\newcommand{\tP}{\tilde{P}}
\newcommand{\tQ}{\tilde{Q}}

\newcommand{\ttt}{\tilde{t}}
\newcommand{\tx}{\tilde{x}}
\newcommand{\tk}{\tilde{k}}

\newcommand{\tF}{\tilde{F}}

\newcommand{\tA}{\tilde{A}}
\newcommand{\tbA}{\tilde{\mathbf{A}}}
\newcommand{\tV}{\tilde{V}}

\newcommand{\Pe}{\mathrm{Pe}\,}

\begin{document}
\title{Dynamics and efficiency of a self-propelled, diffusiophoretic swimmer}
\author{Benedikt \surname{Sabass}}
\author{Udo \surname{Seifert}} 
\affiliation{II. Institut f\"ur Theoretische Physik, Universit\"at Stuttgart,
70550 Stuttgart, Germany}

\begin{abstract}
Active diffusiophoresis - swimming through interaction with a self-generated,
neutral, solute gradient - is a paradigm for autonomous motion at the
micrometer scale. We study this propulsion mechanism within a linear response
theory.
Firstly, we consider several aspects relating to the dynamics of the swimming
particle. 
We extend established analytical formulae to describe small swimmers, which interact with
their environment on a finite lengthscale. Solute convection is also
taken into account. Modeling of the chemical reaction reveals a coupling between
the
angular distribution of reactivity on the swimmer and the concentration field.
This effect, which we term ''reaction induced concentration distortion'',
strongly influences the
particle speed. Building on these insights, we employ irreversible, linear
thermodynamics to formulate
an energy balance. This approach highlights the importance of solute convection
for a consistent treatment of the energetics. The efficiency of swimming
is calculated numerically and approximated analytically. 
Finally, we define an efficiency of transport for swimmers which are moving in
random directions. It is shown that 
this efficiency scales as the inverse of the macroscopic distance over which
transport is to occur.
\end{abstract}
\maketitle

\section{Introduction}
Active motion, driven by an inhomogeneous chemical surface reaction, has
recently attracted much scientific interest. 
A number of different systems, employing this propulsive mechanism on the micro-
and nanometer scale,
have been suggested 
\cite{lammert1996ion,imagilov2001,paxton2005motility,golestanian2005propulsion,
wang2006bipolar, rückner2007chemically, thutupalli2011simple, thakur2011dynamics}.
They open exiting, possible routes for future engineering of nano-devices
\cite{ozin2005dream,paxton2006chemical,He2007,burdick2008synthetic,kaganrapid, solovev2010magnetic, popescu2011pulling}. Furthermore,
self-propelled small
swimmers provide a unique chance to study artificial 
motion at the nanometer scale, e.g., to investigate anomalous diffusion
\cite{golestanian2009anomalous, campos2009superdiffusive, ten2011brownian} or non equilibrium
thermodynamics \cite{palacci2010sedimentation, sano2011}. 
Other interesting questions relating to mutual interactions of these swimmers 
and interactions with the environment are also beginning to be addressed \cite{popescu2009confinement,
yang2010swarm, enculescu2011, thakurinteraction}.

In most of these experiments, a catalytic decomposition of hydrogen peroxide
$H_2O_2$ into oxygen $O_2\,(g)$ and water
is employed for propulsion. In general, these (electro) chemical processes obey
quite
nontrivial kinetics \cite{hall1997electrochemical}. Experiments with bimetallic
nanorods, combining two different metallic catalysts, have shown that the 
mechanism for electrokinetic decomposition of $H_2O_2$ involves an
electrical current inside 
the particle \cite{paxton2006catalytically, wang2006bipolar}. 
Theoretical interpretations of these experiments in terms of a phoretic
propulsion mechanism have recently emerged 
\cite{sundararajan2008catalytic, moran2010locomotion,yariv2010electrokinetic}.
They include driving through an interaction of the swimmer with a 
concentration gradient (diffusiophoresis\cite{anderson1989colloid}) as well as driving through interaction
with a self-generated charge gradient (self-electrophoresis).
Other possible driving mechanisms like interfacial tension
gradients\cite{paxton2004} and nano-bubble formation\cite{gibbs2009autonomously}
have also been investigated.
For a different kind of experiment\cite{howse2007self} polystyrene particles,
half coated with platinum as a catalyst, have been employed. No electric current
inside the swimmer is expected here.
Michaelis-Menten-like kinetics for the chemical reaction were demonstrated and
the data can be consistently explained by a generic
diffusiophoretic/self-electrophoretic model\cite{golestanian2005propulsion}.

Although the understanding of phoretic driving mechanisms has reached an
appreciable level, there remain open questions concerning swimming with 
progressive miniaturization of the swimming particle. One such question is, how
swimming speed changes if the particle size approaches the lengthscale of
solute-swimmer interactions.
Further challenges concern a more detailed, theoretical modelling of the
chemical
surface reaction and the solute convection. Both factors can contribute
to modified predictions for the swimming behavior. All these issues may also be
relevant in terms of optimization for future applications (see citations
above).\\
Finally, the energy balance of the driving mechanism and its efficiency have
only received limited attention. 
Early experimental estimates \cite{paxton2005motility} suggest 
that the efficiency of bimetallic nanorods is very low (on the order of
$10^{-9}$). Here the swimming speed was 
found to be about $1\ldots10\, \mu m/s$. Molecular dynamics simulations of a
swimmer 
in gaseous environment\cite{shi2009computational} yielded much higher
efficiencies of the order $10^{-4}$
and speeds in the order of $m/s$ . The discrepancy emphasizes the
importance of the hydrodynamic dissipation for the 
efficiency. This point found particular emphasize in our 
previous theoretical work were we have examined the hydrodynamic efficiency for
generic surface driven swimmers \cite{sabass2010efficiency}. The hydrodynamic
efficiency provides an upper bound on the overall efficiency while it does
not include
dissipative effects related to building up the chemical gradient driving the
swimmer. 
It is a remarkable property of active phoresis that the overall
efficiency is fully amenable to theoretical calculations. This contrasts with
other active swimming mechanisms,e.g., beating flagella\cite{spagnolie2010optimal}, where some
microscopic, mechanical degrees of freedom are unknown.
Nevertheless, the overall efficiency of a model for 
active, phoretic swimming in a viscous medium has not yet been calculated. 
In the present publication we address the above mentioned 
issues concerning the dynamics and efficiency within a linear response theory
employing the established theoretical framework of neutral
diffusiophoresis\cite{anderson1989colloid}. 

This article is organized as follows. In 
Sec. \ref{sec_model_description}, we introduce the model system and the pertaining equations. 
These are non-dimensionalized and linearized in Sec. \ref{sec_lin_nondim}.
In Sec. \ref{sec_swimming_speed}, we calculate the speed of the swimmer and analyze its
dependence on the interaction between solute and swimmer. The dependence of the swimming speed on 
the chemical surface reaction is investigated in Sec. \ref{sec_react_pol}. In Sec. \ref{sec_energetics}, we  
focus on the energetics of active diffusiophoresis. Here, the efficiency of swimming is analyzed in detail and we 
also consider a measure for the efficiency of transport.

\section{Model description}\label{sec_model_description}
To explore generic features of active diffusiophoresis, we employ an idealized
hydrodynamic model. The swimmer, a spherical particle with radius $R$, 
is placed in an infinitely large container. We assume axial symmetry and use a
spherical 
coordinate system aligned in the
$\mathbf{\hat{e}}_z$ direction. $\tr$ is the distance from the particle
center and $\theta$ is the polar angle with $0\leq\theta\leq\pi$. The unit vectors of the spherical
system are denoted by $ \mathbf{\hat{e}}_r, \,  \mathbf{\hat{e}}_{\theta}$. 
Throughout this publication, we will designate dimensional 
variables with a tilde ( $\tilde{\,}$ ), while constants and non-dimensional
quantities carry no tilde.
\begin{figure}[ht]
\begin{center}
\includegraphics[scale=0.73]{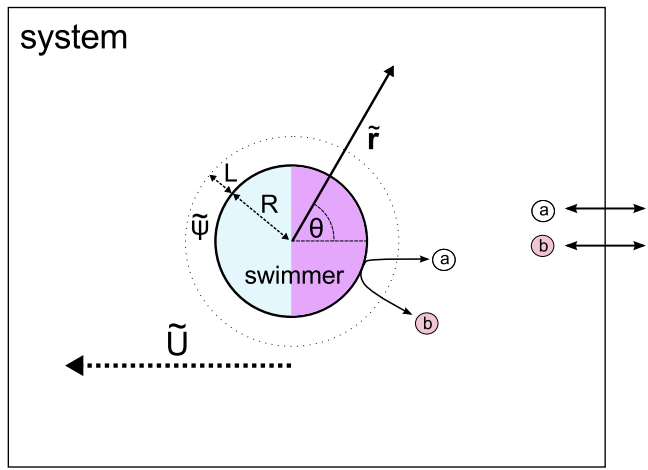}
\caption{\label{fig_1}
Schematic representation of the model for an active swimmer. Molecules of
type $a$ and $b$ are dissolved in the fluid surrounding the swimmer with radius $R$. The molecules 
are transformed into each other through chemical surface reactions and 
the overall concentrations are maintained constant. An interaction between the 
swimmer and the type $b$ molecules occurs through the potential $\tPsi$, whose 
lengthscale is given by $L$. An inhomogeneous distribution of solutes causes the 
swimmer to move with the speed $\tU$. $\mathbf{\tr}$ is the position vector and $\theta$ is the
polar angle in a spherical coordinate system used throughout.
}
\end{center} 
\end{figure}
The multi-component fluid surrounding the swimmer is assumed to be
incompressible and Newtonian. 
It contains two dilute (non-interacting) solutes, whose molecular volumes and
masses are to be similar to that of the solvent. The concentrations of the
solutes are denoted by $\tc_{a}$ and $\tc_{b}$. Solutes and solvent are bound to
interact with the surface of the swimming particle, e.g., through van der Waals,
electrostatic
or steric interactions. In order to simplify the situation, we assume that a
common surface potential  mediates the interactions between both, the solutes,
the solvent and the swimmer. Within our mean-field framework, the common
interaction between swimmer and all fluid constituents merely modifies the
pressure of the fluid. Only for species $b$, we assume, on top of this, a
radially symmetric potential $\tilde{\Psi}$. Physically, this second potential
could, e.g., represent a different polarizability of the solutes $b$. It is
assumed to decay over a characteristic lengthscale $L$. A similar approach has
been taken in a molecular dynamics simulation of the system
\cite{rückner2007chemically}. In Appendix \ref{sec_dilute_solution_Appendix} we
provide a
simple physical model of the three-component solution considered in the
present publication.
Both molecular species, type $a$ and $b$, have the same diffusion constant $D$. 
The concentration fluxes $\tbJ_{\{a,b\}}$ and the corresponding solute
conservation laws are
\begin{align}
\tbJ_{a}& \equiv \tc_{a} \tbv +\tbj_a=\tc_{a} \tbv - D\nabla
\tc_{a},\label{eq_dim_ja}\\
\tbJ_{b}& \equiv \tc_{b} \tbv +\tbj_b=\tc_{b} \tbv - D\left(\nabla \tc_{b}
+\frac{\tc_{b}}{kT}\nabla
\tPsi\right),\label{eq_dim_jb}\\
0&=\nabla \cdot \tbJ_{\{a,b\}}. \label{eq_dim_c_dgl}
\end{align}
The diffusive fluxes of solutes relative to the center of mass are denoted by
$\tbj_{\{a,b\}}$. The center of mass fluid velocity that convects the solutes is
denoted by $\tbv$.
We assume here that the diffusion of solutes near the surface is similar to the
bulk diffusion. An extra surface diffusion term is not taken into account.
Generic boundary conditions,
allowing for an absorption and
emission of solutes at the swimmer's surface, are
\begin{align}
\begin{split}
 \mathbf{\hat{e}}_r\tbJ_{\{a,b\}}(R,\theta)&= \talpha_{\{a,b\}} \, g(\theta),\\
 \nabla\tc_{\{a,b\}}(\infty,\theta)&=0.\label{eq_dim_general_rb}
\end{split}
\end{align}
Here all the $\theta$-dependence of the flux at the surface is contained in the
dimensionless function $g(\theta)$. 
We will always assume that the emission of 
solutes happens in a spatially asymmetric way. $\talpha_{\{a,b\}}$ has the
dimensions of solute flux and quantifies its magnitude.
The Stokes equation and
incompressibility condition read
\begin{align}
\eta\nabla^2 \tbv-\nabla \tp&=\nabla\cdot\tbs =  \tc_b \nabla\tPsi & &\text{and}
& &
 \nabla\cdot \tbv=0.\label{eq_stok_dim}
\end{align}
Here $\eta$ is the viscosity of the fluid, which we assume to be spatially
constant. The pressure accounting for the incompressibility of the fluid is 
denoted by $\tp$ and 
\begin{equation}
 \tbs\equiv \eta \left(\nabla \tbv+(\nabla
\tbv)^{T}\right)-\tp\mathbf{I}= 2 \eta \tilde{\mathbf{E}} -\tp\mathbf{I}
\end{equation}
is the
hydrodynamic stress
tensor. In order to make sure that the total force on the fluid is,
even for a linear concentration gradient ($\tc_b \sim \tr \cos\theta$),
convergent we assume in the following 
\begin{equation}
\tPsi\left(\tr \rightarrow \infty\right) < O\left(\frac{1}{\tr^3}\right).
\end{equation}
The swimming speed is given by $\tU$ in the laboratory frame. In the frame
moving
with the center of the swimmer, it enters the
solution of the Stokes equation through the boundary conditions
\begin{align}
\begin{split}
\tbv(\infty,\theta)&=-\tU \mathbf{\hat{e}}_z,\\
\tbv(R,\theta)&= 0.
\end{split}
\end{align}
The swimming speed $\tU$ is determined from the balance of forces acting on
the swimmer in the overdamped limit
\begin{align}
0=\tF_{\rm m}\hbe_{z} + \int \tc_b \nabla\tPsi \rmd \tilde{V} +
\int \tbs \hbe_r \rmd \tilde{A}_{\tr=R}.\label{eq_force_balance}
\end{align}
Since the swimmer is spherical, we can identify the normal vector on the
surface of 
the sphere with the unit radial vector $\mathbf{\hat{e}}_r$. We are considering
a freely
moving swimmer, therefore the external force $\tF_{\rm m}$ will be set to
zero. The dependence of $\tbs$ on $\tU$ allows the calcuation of the latter
quantity from Eq. (\ref{eq_force_balance}). In a perturbative approach for small
$\tPsi/kT$, one
can assume that the body force $\nabla\cdot\tbs=\tc_b \nabla\tPsi$ is
independent of the fluid velocity $\tbv$. Due to the linearity of the Stokes
equation Eq. (\ref{eq_force_balance}) can be reformulated employing a reciprocal
relation\cite{teubner1982motion}. The result for spherical swimmers is
\begin{equation}
\begin{split}
 0  = \tF_{\rm m} -6 \pi \eta R \tU \,-\int 
[(\frac{3 R}{2 \tr} - \frac{R^3}{2 \tr^3}
-1)\cos\theta\,\mathbf{\hat{e}}_{r}\nabla\cdot\tbs \\
-(\frac{3 R}{4 \tr} + \frac{R^3}{4 \tr^3}
-1)\sin\theta\,\mathbf{\hat{e}}_{\theta}\nabla\cdot\tbs]
\mathrm{d}\tilde{V}
\label{eq_teubner_U}.
\end{split}
\end{equation}
However, when convection strongly modifies the solute distribution $\tc_b$, Eq.
(\ref{eq_teubner_U}) does not permit to calculate $\tU$. 

The swimmer model presented above is only valid if the assumption of a dilute solution
holds.
Corrections to the diffusion coefficient, resulting from mutual solute-solute
interactions, are
proportional to the volume fraction and we hence demand $\tc_{\{a,b\}}\,a^3 \ll
1 $ for solutes with radius $a$. Another source of error is introduced by
the rotational diffusion of the swimmer. The rotating frame results 
in a corrective term for the solute fluxes Eq.
(\ref{eq_dim_ja},\ref{eq_dim_jb}) of the form $-\hbe_{\theta}\tr\, D_{\rm rot}\, \partial_{\theta} \tc_{\rm\{ a,b\}}$
where $D_{\rm rot}$ is the rotational diffusion coefficient of the swimmer. The
relative magnitude
of this correction for the diffusion equations
(\ref{eq_dim_ja}-\ref{eq_dim_c_dgl}) is of the order $R^2 D_{\rm rot}/D$.
Employing the Einstein relations for the diffusion coefficients
$D = kT/\left(6 \pi \eta a\right)$,
$D_{\rm rot}= kT/\left(8 \pi \eta R^3\right)$, we have $R^2 D_{\rm rot}/D = 3
a/(4 R)$. Accordingly, we demand in
the following $a \ll R$. This assumption also justifies disregarding
hydrodynamic 
corrections of the solute-swimmer interaction due to the finite size of the
solutes. 
Taken together, our model is restricted to swimmers which are at least two
magnitudes
larger than the solutes. If the solute radii are in the \AA{} range, we hence
require
$R\,\gtrsim 30\rm\, nm$. 

\section{The linearized, non-dimensional equations}\label{sec_lin_nondim}
Convective transport renders the diffusion equations
(\ref{eq_dim_ja}-\ref{eq_dim_c_dgl})
nonlinear since
the fluid velocity and the concentration fields are both unknown. The fluid
velocity $\tbv$ is determined by Eq. (\ref{eq_stok_dim}) and thus depends itself
on
$\tc_b$. In order to make the system more amenable to analytic calculations we
linearize the non-dimensionalized equations around equilibrium. We thereby
construct a linear response theory for active diffusiophoresis.
The natural length- and energy scales are the radius of the swimmer, $R$, and
the thermal
energy $kT$. Accordingly, we rescale as $r \equiv \tr/R$ and
$\Psi \equiv \tPsi/kT$. The intrinsic lengthscale $L$ of the potential $\Psi$ is
denoted in dimensionless form as 
\begin{equation}
\lambda \equiv L/R. 
\end{equation}
Particle swimming is driven by an asymmetric concentration perturbation of the
equilibrium distribution of $b$-type solutes $\tc^{\rm eq}_{b}(\tr)$. Owing
to the
radial symmetry of $\tPsi$, only the dipole moment of the concentration
perturbation 
\begin{equation}
\frac{3}{2}\int_{-1}^{1} \frac{R\,\talpha_{b}}{D}\, g(\theta)\,
\cos\theta\, \rmd(\cos\theta)
\end{equation}
contributes here. We can accordingly define the 
concentration scale $\hc$, which is relevant for the particle motion, as 
\begin{equation}
\hc \equiv R\,\talpha_{b}/D. \label{eq_hc_def} 
\end{equation}
In the following we will always assume that the concentration perturbation $\hc$
is much smaller than the equilibrium concentration scale. We can then define a 
new dimensionless parameter as
\begin{equation}
 \delta \equiv \hc/\tc^{\rm eq}_b(\infty).\label{eq_delta_def}
\end{equation}
Throughout this publication, we employ $\hc$ for the non-dimensionalization
of concentrations. The equilibrium perturbations of the concentration fields are 
\begin{align}
c_a &\equiv \frac{\tc_a - \tc^{\rm eq}_a(\infty)}{\hc}, \label{eq_c_a_def}\\ 
c_b &\equiv \frac{\tc_b - \tc^{\rm
eq}_b(\infty)e^{-\Psi}}{\hc}\label{eq_c_b_def}.
\end{align}
The solute fluxes $\tbj$ are non-dimensionalizd by $D\hc/R$ and obey the 
boundary conditions 
\begin{align}
&\mathbf{\hat{e}}_r\bj_{\{a,b\}}(1,\theta)=  \frac{R\,\talpha_{\{a,b\}}
\,g(\theta)}{D
\hc} \equiv\alpha_{\{a,b\}} g(\theta). \label{eq_j_bc} 
\end{align}
A typical diffusiophoretic speed magnitude\cite{anderson1989colloid} is given for
$\lambda \lesssim 1$ by
\begin{equation}
\hU \equiv \frac{kT\,L^2}{\eta}
\frac{\hc}{R}. \label{eq_hU_def}
\end{equation} 
We define an associated Pecl\'{e}t number by 
\begin{equation}
 \Pe \equiv \frac{\hU R}{D}. \label{eq_Pe_def} 
\end{equation} Sample calculations, employing measured particle speeds
\cite{paxton2006catalytically, howse2007self}, show that $\Pe$ is typically
small, on the order of $10^{-2}$. The Pecl\'{e}t number thus constitutes,
besides $\delta$, a second small expansion parameter. Employing the
non-dimensional hydrodynamic variables $\bv \equiv \tbv/\hU$ and $p \equiv \tp
R/(\hU \eta )$, the equations for solute conservation
(\ref{eq_dim_ja}-\ref{eq_dim_c_dgl}) and the Stokes equation (\ref{eq_stok_dim})
become
\begin{align}
\nabla^2c_{b}+\nabla\cdot\left(c_{b}\nabla\Psi\right)-\frac{\Pe}{\delta}
\bv\nabla e^{-\Psi}&=\Pe\bv\nabla c_{b},\label{eq_c_b_dgl}\\
\nabla^2c_{a}&=\Pe\bv\nabla c_{a}, \label{eq_c_a_dgl}\\
\nabla^2\bv-\nabla p -  \frac{1}{\lambda^2} c_b\nabla\Psi
&=0.\label{eq_stok_dgl}
\end{align}
The fluid boundary conditions are, as before,
\begin{align}
\begin{split}
  \bv(\infty,\theta)&=-U \mathbf{\hat{e}}_{z},\\
  \bv(1, \theta)&= 0.
\end{split}
\label{eq_v_bc}
\end{align}
The form of Eqns. (\ref{eq_c_b_dgl},\ref{eq_c_a_dgl}) suggests a perturbation
scheme in $\Pe$ to cope with the nonlinearities.  However, difficulties arise if
$\delta$ is of the same magnitude as $\Pe$. 
Then only the right hand sides of Eqns. (\ref{eq_c_b_dgl},\ref{eq_c_a_dgl}) are
small. Furthermore, because $\delta \sim \hc$ and also $\Pe \sim \hc$, it is in
the spirit of
a linear response theory in $\hc$ to set
\begin{align}
\Pe\bv\nabla c_{a} \simeq \Pe\bv\nabla c_{b} \simeq 0
\label{eq_convect_lin_response}.
\end{align}
This linearization approach has the merit that the mobility of the swimmer,
relating $U$ to
$\alpha_{\{a,b\}}$, can be calculated straightforwardly. To do this, one only
needs to
determine the force balance Eq. (\ref{eq_force_balance}) for the two cases
$\alpha_{\{a,b\}}=0$ and $U=0$. This procedure was described in detail by
O'Brien and
White\cite{o1978electrophoretic}. For concrete numerical and analytical
calculations we always rewrite the Stokes equation (\ref{eq_stok_dgl})  by
employing the stream function formalism \cite{happel1983martinus}.

The approximation Eq. (\ref{eq_convect_lin_response}) does not go without any
caveat. The convection terms dominate the diffusion terms in Eqns.
(\ref{eq_c_b_dgl},\ref{eq_c_a_dgl}) 
for large $r$ due to different orders of radial derivatives. In general, one
would therefore follow Acrivos and Taylor \cite{acrivos1962heat} and split the
solution for the
concentration field into an inner solution, with coordinate $r$,
where diffusion dominates and an outer solution, with rescaled radial
coordinate 
$\Pe r$, where convection dominates. We avoid this procedure by 
assuming that $\lambda\, \Pe \ll 1$. Then the interaction
between solutes and swimmers takes place in the diffusion-dominated region
around the swimmer. 

In contrast to the approach taken here, 
the standard way to make analytical progress in view of the
difficulties inherent in Eqns. (\ref{eq_dim_ja}-\ref{eq_stok_dim})  is to assume
that the
magnitude of the surface potential is small ($\Psi(1)\ll 1$). Then convection
does not occur in the $O\left(\Psi(1)^1\right)$ contribution of a regular
perturbation scheme. The resulting  ''Debye-Hueckel'' -like theory is practical
in the case of ionic surface interactions \cite{keh2000diffusiophoretic} but may
become inappropriate for non-ionic interactions when $\Psi(1)$ is not small.
\section{Swimming speed and the role of the interaction potential} \label{sec_swimming_speed}
The nature of the potential $\Psi$, comprising a variety of possible physical
interactions, is important for the diffusiophoretic speed. In order to
demonstrate this in this section, we leave here all issues concerning the
details of the chemical reactions aside. This simplification also facilitates a comparison with
other work \cite{golestanian2005propulsion,howse2007self}. The emission rate of
the solute of type $b$ is in this section given by $g(\theta) =
\left(1+\cos\theta\right)$
and $\talpha_b = \kappa =\mathrm{const.}$ We then have
\begin{equation}
\hbe_r \tbj_b(\theta) = \kappa \left(1+\cos\theta\right)
\label{eq_alpha_const_bc}
\end{equation}
and $\hc = \kappa R/D$. This boundary condition for the surface interacting solute does not contain
$c_a$. Species $a$ is thus irrelevant for the particle swimming. We hence disregard $c_a$ in this section.
The concentration field $c_b$ is determined by the differential Eq.
(\ref{eq_c_b_dgl}) with Eq. (\ref{eq_convect_lin_response}). In spite of its
linearity, there seems to be no analytical solution of this equation available for
arbitrary
$\tPsi$. This is even so in complete absence of convection when $\Pe/\delta=0$. 
\subsection{Analytical approximations}
For short ranged potentials, $\lambda \ll 1$, one can resort to a technique
of matched
asymptotic expansions to calculate the concentration
field\cite{anderson1982motion}. We find for the swimming speed to lowest order
in
$\lambda$, (see Appendix \ref{sec_app_asympt_small_L}),
\begin{equation}
\begin{split}
U_0=\frac{\tU_0}{\hU} = &\lambda^2  \frac{ \kappa
\,kT}{\hU \eta\,D}\frac{R^2}{3}\int_0^{\infty}y\,\left(e^{-\Psi(y)}-1\right)\,\mathrm{d}y = \\
 &\frac{1}{3} \frac{\kappa\,kT\,L^2}{\hU \, D \eta}K_1=\frac{K_1}{3},
\label{eq_U0_smallL_active}  
\end{split}
\end{equation}
where $y\equiv (r-1)/\lambda$ is a dimensionless inner
variable which is $O(1)$ near the surface, where $\Psi$ changes strongly. The
unnormalized moments of the equilibrium excess surface concentration are defined
as
\begin{align}
&K_{\mathrm{n}} \equiv
\int_0^{\infty}y^{\mathrm{n}}\left(e^{-\Psi(y)}-1\right)\,\mathrm{d}y.
\label{eq_Kn_def}
\end{align}
As in the classic electrophoretic Smoluchowski limit
\cite{saville1977electrokinetic}, the dimensional speed $\tU_0$ is
found to be proportional to the
square of the interaction lengthscale ($\sim L^2\,K_1$).
We calculate the first speed correction up to  $O(\lambda)$ for active
diffusiophoresis by
employing methods outlined by Anderson, Lowell and Prieve\cite{anderson1982motion}. The
corresponding first order Pad\'{e} approximant of the swimming speed reads
\begin{equation}
\begin{split}
\tU\approx \tU_0/\left[ 1+ \lambda \left( K_0 + \frac{7\,K_2}{2\,K_1} +
\frac{\Pe}{\delta}\frac{M}{2} +\frac{N}{K_1}\right) \right],
\end{split}
\label{eq_active_U_smol}
\end{equation}
where the following definitions are employed  
\begin{align}
M \equiv & \int_0^{\infty}
\int_0^{y}y'\,\left(e^{-\Psi(y')}-1\right)\,\mathrm{d}y'
\left(e^{-\Psi(y)}-1\right)\,\mathrm{d}y,\label{eq_M_def}\\
\begin{split}
 N \equiv & -2
\int_0^{\infty}\int_0^{y}y'\,\left(e^{-\Psi(y')}-1\right)\,\mathrm{d}y'
\left(e^{\Psi(y)}-1\right)\,\mathrm{d}y\\
&+\int_0^{\infty}y^{2}\left(e^{-\Psi(y)}-1\right)\left(e^{\Psi(y)}-1\right)\,
\mathrm{d}y.\label{eq_N_def}
 \end{split}
\end{align}
Note that the chemical reaction must occur at a finite
distance from the physical hard-core boundary of the swimmer. Therefore, the
potential $\Psi(r=1)$ cannot diverge in 
active diffusiophoresis and all the constants defined above remain finite.
\subsection{Numerical results}
\begin{figure}[ht]
\begin{center}
\includegraphics[scale=0.87]{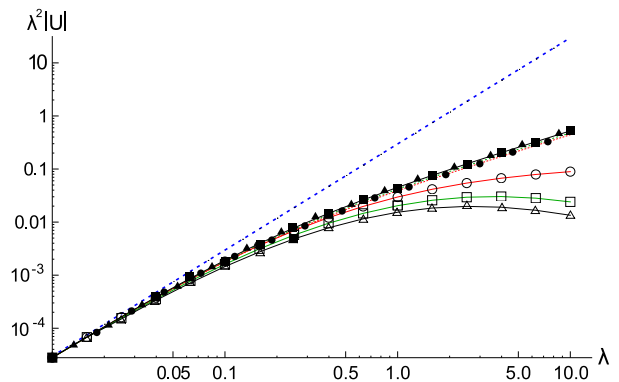}
\caption{\label{fig_2-1}
Swimming speed vs. interaction lengthscale $\lambda$ for an exponential
repulsion:
$\Psi(r) = \exp\left[-\left(r-1\right)/\lambda \right]$.
~(Dashed line) Swimming speed in lowest order approximation $\lambda^2|U_0|$
for $\lambda \ll 1$ from Eq. (\ref{eq_U0_smallL_active}).
~($\circ$) Numerical solution for $\Pe/\delta=0$.
~($\bullet$) Pad\'e approximant, Eq. (\ref{eq_active_U_smol}), for
$\Pe/\delta=0$.
~($\square$) Numerical solution for $\Pe/\delta=5$.
~($\blacksquare$) Pad\'e approximant for $\Pe/\delta=5$.
~($\triangle$) Numerical solution for $\Pe/\delta=10$.
~($\blacktriangle$) Pad\'e approximant for $\Pe/\delta=10$.}
\end{center} 
\end{figure}

\begin{figure}[ht]
\begin{center}
\includegraphics[scale=0.87]{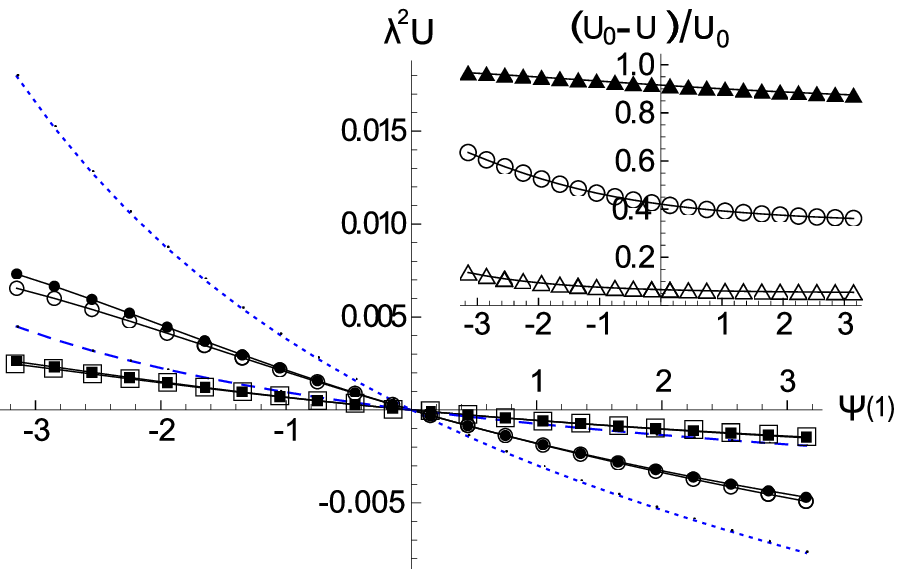}
\caption{\label{fig_2-2}
Swimming speed vs. interaction potential magnitude $\Psi(1)$ for
$\Psi(r) = \Psi(1) \exp\left[-\left(r-1\right)/\lambda \right]$ with $\Pe/\delta
=0$.
~($\square$) Numerical solution for $\lambda = 0.05$.
~($\blacksquare$) Pad\'e approximant, Eq. (\ref{eq_active_U_smol}), for
$\lambda = 0.05$.
~(Dashed line) $\lambda^2|U_0|$ from Eq. (\ref{eq_U0_smallL_active}) for
$\lambda =
0.05$.
~($\circ$) Numerical solution for $\lambda = 0.1$.
~($\bullet$) Pad\'e approximant for $\lambda
= 0.1$.
~(Dotted line) $\lambda^2|U_0|$ for $\lambda =0.1$.
Inset: Relative deviation of the lowest order approximation
$\left(U_0-U\right)/U_0$ vs. $\Psi(1)$ with $\Pe/\delta =0$.
~($\blacktriangle$)  $\lambda = 1$.
~($\circ$) $\lambda = 0.1$.
~($\vartriangle$) $\lambda = 0.01$.
}
\end{center} 
\end{figure}

\begin{figure}[ht]
\begin{center}
\includegraphics[scale=0.87]{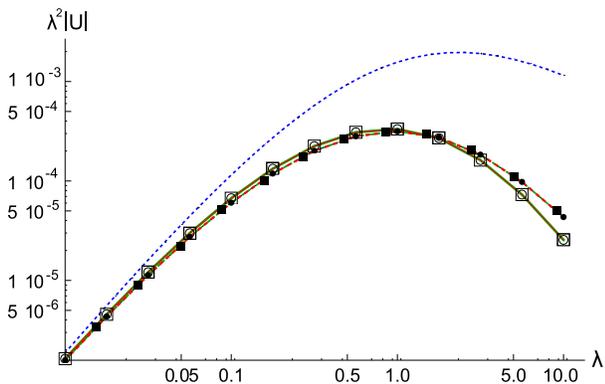}
\caption{\label{fig_2-3}
Swimming speed vs. $\lambda$ for a van der Waals -like attraction: 
$\Psi(r) = -A \lambda^3 /(r-1+\lambda)^3 /(r-1+\lambda+2)^3$ with
$A = 1$. $\lambda$ is here interpreted as solute radius divided by
the swimmer radius and the potential is truncated at a distance 
$\lambda$ away from the surface.
~(Dashed line) $\lambda^2 |U_0|$ from Eq. (\ref{eq_U0_smallL_active}).
~($\circ$) Numerical solution for $\Pe/\delta=0$.
~($\bullet$) Pad\'e approximant, Eq. (\ref{eq_active_U_smol}), for
$\Pe/\delta=0$.
~($\square$) Numerical solution for $\Pe/\delta=5$.
~($\blacksquare$) Pad\'e approximant for $\Pe/\delta=5$.}
\end{center} 
\end{figure}

In order to go beyond the $\lambda \ll1$ limit, we solve Eqns.
(\ref{eq_c_b_dgl}-\ref{eq_stok_dgl},\ref{eq_convect_lin_response})
numerically for the boundary conditions given in Eqns.
(\ref{eq_v_bc},\ref{eq_alpha_const_bc}). Results are displayed in Figs.
\ref{fig_2-1}, \ref{fig_2-2} and \ref{fig_2-3} where the swimming speed is
plotted 
as $\lambda^2 U = L^2 \tU/(R^2 \hU )$ in order to demonstrate the physical dependence of the speed 
on the interaction length $L$. As it is the case for other phoretic
effects\cite{saville1977electrokinetic,prieve1987diffusiophoresis}, convection
causes a non-monotoneous 
relation between particle speed and $\lambda$ (Fig. \ref{fig_2-1}). An
increasing magnitude
of the 
surface potential $\Psi(1)$ reduces the range of validity of the lowest order approximation $U_0$ to smaller
values of 
$\lambda$ (Fig. \ref{fig_2-2}). The Pad\'e approximant of $U$, Eq.
(\ref{eq_active_U_smol}), 
then describes the case of strong surface interactions much more satisfyingly
\cite{anderson1991diffusiophoresis}. 
Therefore, the Pad\'e approximant presents a significant improvement over the
lowest order estimate
$U_0$,  in spite of being useful only for $\lambda < 1$. 

Fig. \ref{fig_2-3} shows the swimming speed for a truncated van der Waals-like
interaction where the potential decays $\sim 1/r^6$ 
only far away of the swimmer. The result demonstrates that multiple 
lengthscales of the interaction potential may modify the simple
scaling of $\tU_0 \sim L^2$. Furthermore, the similarity of the results 
for $\Pe/\delta=0$ and $\Pe/\delta=5$ qualifies the simple 
notion that convection of solutes reduces the speed of the particle. Since, for Fig. \ref{fig_2-3},
we have $\Psi(r) \ll 1$ when $\lambda > 1$,  the influence of the Pecl\'et number
is suppressed almost completely. This dependence of the effect of convection on the
strength of the surface interaction also appears through the constant $M$ in the Pad\'e approximant given in
Eqns. (\ref{eq_active_U_smol},\ref{eq_M_def}). 
\subsection{Comparison with passive diffusiophoresis}
It is of interest to compare the swimming speed given in Eqns.
(\ref{eq_U0_smallL_active},\ref{eq_active_U_smol}) with the analogous formulae
for passive diffusiophoresis in an externally imposed concentration gradient
\cite{anderson1982motion}. Here the boundary conditions for the concentration of
the solute of type $b$ read
\begin{align}
\begin{split}
\mathbf{\hat{e}}_r\tbj_{b}(1,\theta)&=0,\\
\nabla \tc|_{\infty}&=\text{const.}\times\mathbf{\hat{e}}_z.
\end{split}
\end{align}
The lowest order result for swimming speed in passive diffusiophoresis $\tU^{p}$
is 
\begin{equation}
 \tU_0^{p} = |\nabla \tc|_{\infty} \frac{kT 
L^2}{\eta}\,K_1 \label{eq_U0_smallL_passive}.
\end{equation}
The formulae (\ref{eq_U0_smallL_active}) and (\ref{eq_U0_smallL_passive}) for
active
and passive swimming are, apart from a replacement of $\kappa/3D$ by $|\nabla
\tc|_{\infty}$, the same. These different prefactors result from the dipole
moments of the concentration fields around the swimmer. The concentration far
away from the
particle is given by $\tc_b\approx R\kappa/D\left[1/r +\cos\theta /(2r^2)\right]$
in our case and
$\tc_b\approx R |\nabla \tc|_{\infty}\left[r +1/(2r^2)\right]\cos\theta$ for an
externally imposed concentration gradient.
At $r=1$, the self generated dipole is therefore only $\sim \kappa R D/2$ 
while the imposed concentration dipole is $\sim 3 \,R |\nabla
\tc|_{\infty}/2$. This leads to the differing factor of $1/3$, appearing in the swimming speeds. 

The general agreement of both lowest order formulae for active and passive
diffusiophoresis can be rationalized by noting that both concentrations are, 
in close proximity to the swimmer's surface, in radial equilibrium.
Accordingly, the lowest order solute flux also vanishes for 
active diffusiophoresis since the
diffusive exchange of solute near the active swimmer is $\sim\partial
c_b/\partial(\lambda
y)=O(1/\lambda)$ while the emission/absorption of solute, determined by $g(\theta)$, is
only of $O(1)$. 

For passive diffusiophoresis\cite{anderson1991diffusiophoresis}, the analogous formula to the first order swimming
speed Eq. (\ref{eq_active_U_smol}) is, in our notation,
\begin{equation}
\tU^{p}\approx \tU_0^{p}/\left[ 1 + \lambda \left( K_0 +
\frac{K_2}{2\,K_1} + \frac{\Pe}{\delta}\frac{M}{2} \right) \right]
\label{eq_passive_U_smol}.
\end{equation}
The main difference between (\ref{eq_active_U_smol}) and
(\ref{eq_passive_U_smol})
is the appearance of the term $\lambda N/K_1$ for active diffusiophoresis. The
constant $N$, defined in Eq. (\ref{eq_N_def}), contains an integral over
$\exp{\left(\Psi(r)\right)}$. It increases the 
relative importance of the $\lambda^1$ correction for strong repulsive surface
interactions. For monotonic potentials, $N$ is a positive quantity. Therefore, 
the correction for active swimmers increases the swimming speed for
purely repulsive interactions where we have $K_1 <0$.
Convective corrections $\sim\Pe/\delta$ are, both for active and passive swimming, relevant if the
equilibrium concentration scale $\tc^{\rm eq}_b$ is much
larger than the concentration disturbance $\hc$. It is an interesting side note
to the convective correction that the constant
$M$ is related to the hydrodynamic dissipation in the 
boundary layer\cite{sabass2010efficiency} $\tilde{W}_{\rm hyd} \approx L^3\,M\,4\pi\, 
\left(kT\, \hc\right)^2/\left(3 \eta\right)$.

\section{Reaction induced concentration distortion}\label{sec_react_pol}
In general, the kinetics of the chemical transformations at the swimmer's
surface depend on the local concentrations and correlation functions of educts
and products. An inhomogeneous reactivity profile on the swimmer, realized,
e.g., through partial coating with a catalyst, couples to the angular solute
distributions. Therefore, the resulting solute flux at the swimmer will
usually not have the
same angular dependence as the reactivity profile. We term this effect 
''reaction induced concentration distortion'' and it bears a certain analogy to 
polarization phenomena, inasmuch as it has a possibly nonlinear effect on the
swimming speed. 
\subsection{Boundary conditions for the concentration field}
For the considered case of dilute solutions one might anticipate that the
fraction of unoccupied catalytic sites on the swimmer is always very small. Then
the overall reaction rates depend linearly on the concentrations of the
reactants. Concentration fields resulting from nonlinear
reactions\cite{park2008concentration} 
are beyond the scope of this article. We therefore 
assume, for simplicity, a reversible, first order reaction of the form $a
\leftrightarrow b$. For the remainder of this article we will employ
as boundary conditions for the concentration fields at $r =1$
\begin{align}
\mathbf{\hat{e}}_r\tbj_{b}(1,\theta)& =
\left(1+\cos\theta\right)\left[\tk_{\rm ab}\tc_a(1,\theta)-\tk_{\rm
ba}\tc_b(1,\theta)
\right]
\label{eq_bc_mass_action_b},\\
\mathbf{\hat{e}}_r\tbj_{a}(1,\theta)& =
-\mathbf{\hat{e}}_r\tbj_{b}(1,\theta) \label{eq_bc_mass_action_a},
\end{align}
where $\tk_{\rm ab}$ and $\tk_{\rm ba}$ are rate constants which we
non-dimensionalize with $D/R$. The concentration dependence of the flux at the
swimmer surface is also termed a radiation boundary condition. Similar
mathematical problems have occurred in the calculation of mean
first passage times for the combined encounter and reaction of asymmetric
molecules \cite{solc1973kinetics, shoup1981diffusion}. 

In equilibrium, where the fluxes vanish, the radially symmetric
distributions of solutes obey $\tc_{a}^{\rm eq}(r) =\tc_{a}^{\rm
eq}(\infty)$,
$\tc_{b}^{\rm eq}(r) =\tc_{b}^{\rm eq}(\infty) e^{-\Psi(r)}$ and 
\begin{equation}
 \frac{\tc_{a}^{\rm eq}\left(\infty\right)}{\tc_{b}^{\rm eq}\left(\infty\right)}
= \frac{\tk_{\rm ba} e^{-\Psi(1)}}{\tk_{\rm ab}}.
\end{equation}
A finite reaction rate at the swimmer's surface is ultimately driven by chemical
potential
differences far away, at 
the boundary of the system. Expressions for the chemical potentials are given in
Appendix \ref{sec_dilute_solution_Appendix}. 
Within the linear response regime we expect that
~$\left(\tc_{\{a,b\}}(\infty) /\tc_{\{a,b\}}^{\rm eq}(\infty)-1\right) \ll 1$.
Therefore, the chemical
potential difference at $\tr \rightarrow \infty$ becomes
\begin{equation}
\begin{split}
\Delta \mu_{\infty} \equiv \frac{\tmu_a(\infty)- \tmu_b(\infty)}{kT} \approx
\frac{\tc_a\left(\infty\right)}{\tc_a^{\rm eq}(\infty)}
-\frac{\tc_b\left(\infty\right)}{\tc_b^{\rm eq}(\infty)} =\\
\frac{1}{\tk_{\rm ab} \tc_{a}^{\rm eq}}\left[\tk_{\rm
ab}\,\tc_a(\infty)-\tk_{\rm ba}e^{-\Psi(1)}\,\tc_b(\infty)
\right]. 
\end{split}
\end{equation}
The concentration scale Eq. (\ref{eq_hc_def}) can now be defined as
\begin{equation}
\hc =  \frac{R}{D}\;\tk_{\rm ab}\,\tc_a^{\rm eq}(\infty) \,  \Delta
\mu_{\infty}=k_{\rm ba}\,\tc_b^{\rm eq}(\infty)e^{-\Psi(1)} \, \Delta
\mu_{\infty} \label{eq_hc_active_def}
\end{equation}
and from Eq. (\ref{eq_dim_general_rb}) with Eq. (\ref{eq_bc_mass_action_b}) we
have $g(\theta)= \mathbf{\hat{e}}_r\tbj_{b}(1,\theta)\times R/(D\,\hc)$.
The parameter $\delta$ and the velocity scale $\hU$ are accordingly given by
\begin{align}
\delta &= \frac{\hc}{\tc_{b}^{\rm eq}(\infty)} = \Delta
\mu_{\infty}e^{-\Psi(1)} \, k_{\rm ba},
\label{eq_delta_active_def}\\
 \hU &= \frac{kT L^2}{\eta} \frac{ k_{\rm ab} \tc^{\rm eq}_a \Delta
\mu_{\infty}}{R}. \label{eq_Uscale_def}
\end{align}

In order to calculate the concentration perturbations $c_{\{a,b\}}$ around the
swimmer we proceed by expanding them in Legendre polynomials
$P_n(\cos\theta)$ as
\begin{align}
\begin{split}
 c_{a} = &c_a\left(\infty\right)+\sum_{n=0}^{\infty} c_{a}^{n}(r)
P_n\left(\cos\theta\right),\\
c_{b} = &c_b\left(\infty\right) e^{-\Psi(r)} + \sum_{n=0}^{\infty} c_{b}^{n}(r)
P_n\left(\cos\theta\right)\label{eq_c_LegPol_expansion},
\end{split}
\end{align}
which includes the boundary conditions at $r \rightarrow \infty$.
The boundary conditions at the surface of the swimmer, Eqns.
(\ref{eq_bc_mass_action_b},\ref{eq_bc_mass_action_a}), couple
different coefficients of the expansion Eq. (\ref{eq_c_LegPol_expansion}) and
one has
\begin{align}
\begin{split}
\frac{-2}{2n+1}\left[\p_r c_{b}^n + c_{b}^n\p_r\Psi\right]|_{r=1} =&
\int_0^{\pi}P_n\left(\cos\theta\right) g(\theta) \sin\theta
\mathrm{d}\theta, \label{eq_bc_a_react_pol}
\end{split}\\
-\left[\p_r c_{b}^n + c_{b}^n\p_r\Psi\right]|_{r=1} =&\left[\p_r c_{a}^n
\right]|_{r=1}. \label{eq_bc_b_react_pol}
\end{align}
\subsection{Analytical approximations}
We consider a swimmer with short interaction length $\lambda\ll 1$. Employing
the methods presented in
Appendix \ref{sec_app_asympt_small_L} we calculate the concentration fields to
leading
order in $\lambda$. 
Far away of the swimmer, where $\Psi(r)\rightarrow 0$, Eqns. (\ref{eq_c_b_dgl},\ref{eq_convect_lin_response})
can be replaced by Laplace's equations and we have $c_{\{a,b\}}^n(r)=
A^n_{\{a,b\}}/r^{n+1}$ in Eq.
(\ref{eq_c_LegPol_expansion}). For $c_a$, this expansion is valid throughout the
whole system. For the $b$-type solute we have to leading order near the surface of
the swimmer $c_{b}^n(y) \approx   a_{b}^n e^{-\Psi(y)}$. Matching these
solutions and employing the boundary
condition Eq. (\ref{eq_bc_b_react_pol}) leads to $a_b^n=A_b^n=-A_a^n$. Finally,
Eq. (\ref{eq_bc_a_react_pol}) yields a recursion equation for the constants
$A_{b}^n$ 
\begin{equation}
\begin{split}     
& \left[2\delta_{n,0}+\frac{2}{3}\delta_{n,1}\right]  =\\
&\frac{2\left(n+1+k_+\right)}{2n+1}A_b^n+\frac{2n\,k_+}{
\left(2n+1\right)\left(2n-1\right)}A_b^{n-1}+\\
&\frac{2 (n+1)k_+}{\left(2n+1\right)\left(2n+3\right)}
A_b^{n+1}  \label{eq_recurs_Ab}.
\end{split}
\end{equation}
Here we defined
\begin{equation}
 k_+\equiv k_{\rm ab}+k_{\rm ba} e^{-\Psi(1)} = k_{\rm
ab}\left(1+\frac{\tc_{a}^{\rm eq}\left(\infty\right)}{\tc_{b}^{\rm
eq}\left(\infty\right)}\right). \label{eq_kp_def}
\end{equation}
Eq.(\ref{eq_recurs_Ab}) can be written in matrix form as $B_j=M_{jn}\,A_b^{n}$.
The
off-diagonal elements of $\{M_{jn}\}$
decay like $\sim 1/n$ and we can invert 
a finite matrix $\{M_{jn}\}$ to determine a numerical approximation of the
$\{A_b^{n}\}$. In the following, plots of analytical results involving $A^n_b$,
are 
created by employing Eq. (\ref{eq_recurs_Ab}) and $n_{\rm max} = 70$.
Since $\Psi(r)$ is radially symmetric, the
swimming speed depends only on the dipole moment of the concentration field.
Accordingly, an
expansion of Eq. (\ref{eq_teubner_U}) for small $\lambda$ yields for the lowest
order the free swimming speed
\begin{equation}
\tU_0 = \hU \frac{K_1}{3} \, 2 A^1_b.
\label{eq_U0_smallL_active_react_pol}  
\end{equation}
For small bare rates, and therefore small $k_+$, one can expand the inner
concentration field of the $b$-type solute  
\begin{equation}
\begin{split}
  c_b =& c_b(\infty) e^{-\Psi(r)} + e^{-\Psi(r)} \times\\
       & \left[A^0_b+A^1_b\cos\theta+ A^2_b P_2(\cos\theta) + A^3_b
P_3(\cos\theta)+O(k_{+}^3) \right]
\end{split}
\end{equation}
with following approximations for the constants
\begin{align}
 A_b^0 &\approx  \frac{42 + 2 k_+}{42 + 51 k_+}, & & A_b^1\approx  \frac{135 -
26 k_+}{270 + 353 k_+},\\
 A_b^2 &\approx  -\frac{2k_+}{18 + 33 k_+}, & & A_b^3\approx \frac{k_+^2}{60}.
\end{align}
The magnitude of the dipole moment is determined by $A^1_b \approx \left(1/2 - 3
k_+/4\right)$. Reaction induced concentration distortion, emerging here through 
the corrections in orders of $k_+$, reduces the dipole 
moment and thus slows the particle swimming down.
\subsection{Numerical results}
\begin{figure}[ht]
\begin{center}
\includegraphics[scale=0.87]{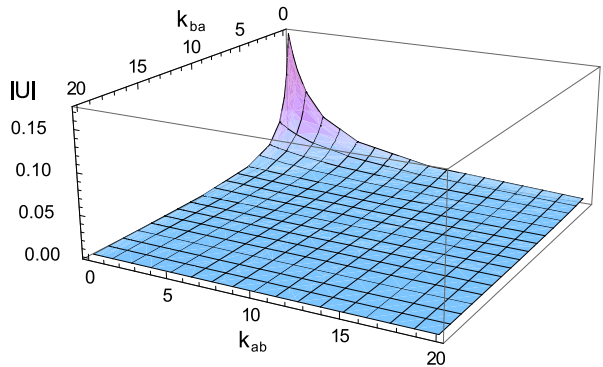}
\caption{\label{fig_3-1} Particle speed $U$ as a function of 
$k_{\rm ab}$ and $k_{\rm ba}$ for ~$\Psi(r) =
\exp\left[-\left(r-1\right)/\lambda \right]$,
 $\Pe/\delta=1$ and $\lambda = 0.1$.}
\end{center} 
\end{figure}

To complement the analytical approximations, we calculate the swimming 
speed numerically. Eqns. (\ref{eq_bc_mass_action_b},\ref{eq_bc_mass_action_a})
with fixed concentrations far away of the swimmer are employed
for numerical solution of Eqns.
(\ref{eq_c_b_dgl}-\ref{eq_convect_lin_response}).
The boundary conditions result in an infinite system of equations where the solute
concentrations far away of the swimmer $c_{\{a,b\}}(\infty)$
determine the reaction
speed. The system can be truncated above a certain order $n_{\rm max}$ of the
Legendre polynomials.
In choosing here $n_{\rm max}=4$ we made sure that the error in the calculated concentrations is negligible. Fig. \ref{fig_3-1} contains an exemplary plot
of how the reaction induced concentration distortion influences the swimming speed $U$. 
The chosen velocity scale $\hU$ (Eq. \ref{eq_Uscale_def}) contains the linear dependence on 
$c_{\{a,b\}}(\infty)$. 
With the employed first order reactions, the concentration distortion depends 
nonlinearly on the bare rates $k_{\rm ab}$ and $k_{\rm
ba}$ but linearly
on the concentration scales $c_{\{\rm{a,b}\}}(\infty)$. Thus, 
in an experiment in the linear regime, with fixed $k_{\rm ab}$ and $k_{\rm ba}$,
the reaction induced concentration distortion might be accounted for by a constant prefactor, modifying the
swimming
speed. 
\begin{figure}[ht]
\begin{center}
\includegraphics[scale=0.87]{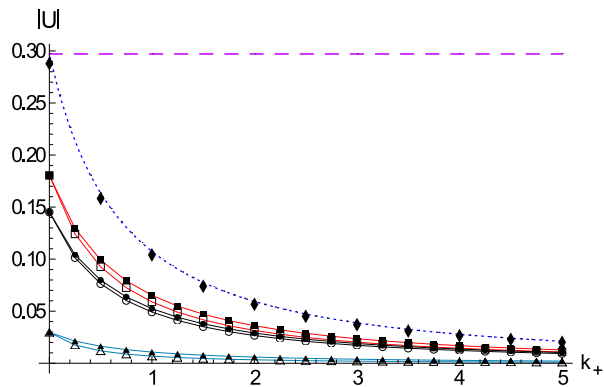}
\caption{\label{fig_3-2}Rate constant dependence of the swimming speed $U$ for $\Psi(r)=\exp\left[-\left(r-1\right)/\lambda \right]$.
 See Eq. (\ref{eq_kp_def}) for the definition of $k_+$.
(Dashed line) Reaction induced concentration distortion neglected as explained in the text, leading to $U = K_1/3$ with $K_1 \simeq -0.89$.
(Dotted line) Approximation $U_0$ (Eq. (\ref{eq_U0_smallL_active_react_pol})) for
$\lambda \ll 1$ which takes the concentration distortion into account.
Full symbols are numerical results for $k_{\rm ba} =0$ and open symbols are
results for $k_{\rm ab} =0$.
($\blacklozenge$) $\lambda=0.005$, $\Pe/\delta =0$.
($\blacksquare$,$\square$) $\lambda=0.1$, $\Pe/\delta =0$.
($\bullet$,$\circ$) $\lambda=0.1$, $\Pe / \delta =10$.
($\blacktriangle$,$\vartriangle$) $\lambda=1$, $\Pe / \delta =0$.}
\end{center} 
\end{figure}

In Fig. \ref{fig_3-2} we plot numerical results for the swimming speed as a function of $k_+$, defined in Eq. (\ref{eq_kp_def}).
Neglecting the reaction induced concentration distortion, the
naive boundary condition for the concentration would be $\hbe_r \bj_b|_{r=1}
=\left(1+\cos\theta \right)$. The resulting swimming speed is independent of $k_+$.
It agrees with the analytical approximation in Eq. (\ref{eq_U0_smallL_active_react_pol}) 
in the limit $k_+ \rightarrow 0$. However, Fig. \ref{fig_3-2} shows that neglecting reaction 
induced concentration distortion in this way leads to significant errors in the speed estimate for
finite reaction rate constants. The analytical approximation, Eq. (\ref{eq_U0_smallL_active_react_pol}), 
is found to be useful for $\lambda \lesssim 0.01$.

\begin{figure}[ht]
\begin{center}
\includegraphics[scale=0.87]{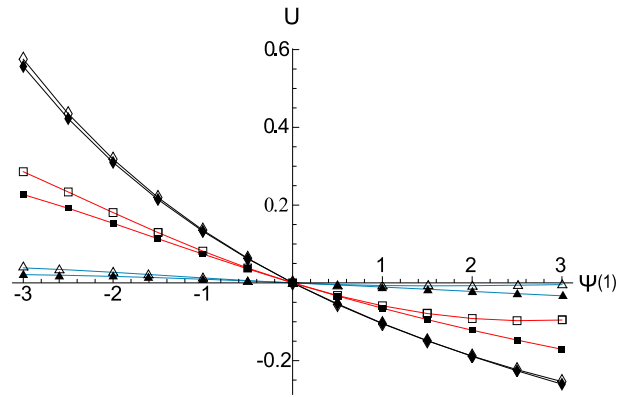}
\caption{\label{fig_3-3}Swimming speed as a function of the potential strength $\Psi(1)$. 
Numerical results are computed with $\Psi(r) = \Psi(1)
\exp\left[-\left(r-1\right)/\lambda \right]$ and  $\Pe/\delta =0$.
Full symbols indicate $(k_{\rm ab} =1,\,k_{\rm ba} \exp\left(\Psi(1)\right)=0$).
Open symbols indicate $(k_{\rm ab} =0,\,k_{\rm ba}
\exp\left(\Psi(1)\right)=1$). 
($\blacklozenge$,$\lozenge$) $\lambda=0.005$.
($\blacksquare$,$\square$) $\lambda=0.1$.
($\blacktriangle$,$\vartriangle$) $\lambda=1$.
}
\end{center} 
\end{figure}

Fig. \ref{fig_3-3} shows the dependence of swimming speed on the strength of the interaction potential $\Psi(1)$. 
The symmetry between the effects of changing $k_{\rm ab}
\exp\left(-\Psi(1)\right)$ and $k_{\rm ba}$, apparent in the parameter $k_+$ in $U_0$ when $\lambda \ll 1$, is lost for
finite $\lambda$. Increasing the strength of the
interaction potential $\Psi(1)$ makes this asymmetry more pronounced.
\section{Energetics}\label{sec_energetics}
\subsection{Energy balance}\label{sub_sec_energy_balance}
We consider an isothermal situation ($T = \mathrm{const.}$) where 
the system, i.e., the swimmer and the multi-component fluid is in a steady
state. The steady state implies, that the molecules which are modified by a chemical reaction
need
to be replenished from outside the system. To do this in a real experiment, one
would need 
to connect some sort of external apparatus to the system. We idealize the
apparatus by a 
reversible process consuming the power $\tP_{\rm in}$. The total of external
apparatus
and system does not exchange matter with the external world. Therefore
$\tP_{\text{in}}$ 
is balanced by the overall heat outflow from the system and apparatus. On
employing classical, linear nonequilibrium thermodynamics we find the power
input (see Appendix \ref{sec_app_energy_balance})
\begin{equation}
\begin{split}
 \tP_{\rm in} = kT\, \hc D R \int \left( \Theta_{\rm hyd}+ \Theta_{\rm diff} + 
\Theta_{\rm react} \right) \,\rmd V \label{eq_pin_entropy_prod}.
\end{split}
\end{equation}
The power is consumed by the following three entropy production rates per volume
$\Theta_{(\ldots)}$
\begin{align}
&\Theta_{\rm hyd} \equiv -\Pe \bv\,\left( c_b +\frac{e^{-\Psi}}{\delta}  \right)\nabla \Psi,\\
&\Theta_{\rm diff} \equiv -\bj_b \nabla\Psi - \sum_i \bj_i \nabla  \mu_i,\\
&\Theta_{\rm react} \equiv  \delta \left( r-1 \right) g(\theta)
\left(\mu_a-\Psi -\mu_b\right).
\end{align}
The integral over $\Theta_{\rm hyd}$ is the overall hydrodynamic work 
because no external force is applied to the swimmer and because
$\nabla\cdot\bv=0$. 
Its proportionality to $\Pe$ emphasizes an interpretation as convection of
$b$-type
solute within the potential $\Psi$. Furthermore, convection of solutes also
modifies 
the reaction rate through the mass action law underlying the definition of 
$g(\theta)$. Hence, the common neglect of convection in dynamical
problems\cite{anderson1989colloid} would make
an overall energy balance within the chosen framework impossible. 
In the linear response regime the power consumption up to $O(\delta^2)$ can 
be rewritten for the employed model as (see Appendix \ref{sec_app_energy_balance})
\begin{equation}
\begin{split}
\tP_{\rm in}= kT D \hc R\,\int \bJ_b \Delta\mu_{\infty} \hbe_r
\mathrm{d}A_{r=\infty}. \label{eq_pin_model}
\end{split}
\end{equation}
Note that $\Delta\mu_{\infty} $ is the chemical potential difference at the
outer boundary of the system, not 
the local chemical potential difference at the the reaction site. Eq.
(\ref{eq_pin_model}) is nothing but
the free energy exchange between the system and the apparatus. Due to the
similarity of masses and sizes of the fluid 
constituents, it does not make a difference whether we consider a fixed pressure
at the outer boundaries of the system or a fixed system size.

\subsection{Efficiency of swimming}
Our freely moving swimmer does not have an external power output
which could be employed to calculate its efficiency. Still, one might ask how
efficient
this swimmer can transport itself. A natural way to do this, is to compare the
energy dissipation of active swimming with the energy dissipation taking place
when
dragging the same particle. We accordingly define a swimming efficiency as
\begin{equation}
\begin{split}
\epsilon \equiv &\frac{6 \pi \eta R \,\tU^2}{ \tP_{\rm in}} = \Pe \lambda^2
\frac{6\pi\,U^2}{ \int
 \bJ_b \Delta\mu_{\infty} \,\hbe_r \rmd A_{r=\infty}} =\\
  &\frac{\Pe}{\delta}  \, k_{\rm ba} e^{-\Psi(1)} \lambda^2 \frac{6\pi\,U^2}{
\int
 \bJ_b \hbe_r\rmd A_{r=\infty}}\label{eq_eff},
\end{split}
\end{equation}
where we have used Eq. (\ref{eq_delta_active_def}) in the second line. 
The numerator of Eq. (\ref{eq_eff}) is the hydrodynamic dissipation of a
passively
dragged sphere and
the denominator is the power consumption of our external apparatus, providing
the
energy for active swimming. This definition is a natural extension to
Lighthill's formula for
hydrodynamic efficiency \cite{lighthill1952squirming}. The power consumption
$\tP_{\rm in}$
is bounded from below by the hydrodynamic dissipation
$2 \eta \int\mathbf{\tE}:\nabla\tbv \, \rmd \tV$.
This intuitive result follows formally in Eq. (\ref{eq_pin_entropy_prod}) from
the
linear force-flux relationships. Since
\cite{sabass2010efficiency} $6 \pi \eta R \,\tU^2 \leq 2 \eta \int
\mathbf{\tE}:\nabla\tbv \, \rmd
\tV$, we always have $\epsilon \leq 1$ as long as no approximation is used to
evaluate Eq. (\ref{eq_eff}).
\subsection{Analytical approximation for the efficiency of swimming}
In the limit $\lambda \ll 1$ one can employ the lowest order speed
$\tU_0$, Eq. (\ref{eq_U0_smallL_active_react_pol}), for the calculation of the
efficiency
from Eq. (\ref{eq_eff}). The result, including the effect of reaction induced concentration distortion, is
\begin{equation}
\begin{split}
\epsilon \approx &\lambda^2 \;\Pe \frac{6 \pi}{4\pi\,A_b^0 \Delta\mu_{\infty}}
\left(\frac{2}{3}A_b^1 K_1\right)^2 = \\
&\frac{D_S}{D}\;  \frac{L^4 \,\hc}{R}\;\frac{kT}{\Delta\tmu_{\infty}}
\;\frac{4\pi\,\left(A_b^1 K_1\right)^2}{A_b^0} \label{eq_eff_asympt},
\end{split}
\end{equation}
where we have employed the translational diffusion constant of the swimmer 
$D_{\rm S} = kT/6\pi\eta R$. The first three dimensionless groups in the second
line of Eq. (\ref{eq_eff_asympt}) determine
the magnitude of the efficiency. As discussed in our previous
work\cite{sabass2010efficiency}, (nano-) swimmers with interaction lengths 
comparable to their size can have a higher efficiency than swimmers with $L \ll R$. This
is is evident from the factors $D_{\rm S}/D$ and $\hc \,L^4/R = \hc L^3\times
\lambda$
in Eq. (\ref{eq_eff_asympt}). We evaluate Eq. (\ref{eq_eff_asympt}) by employing
the parameters $A_b^0, A_b^1$ calculated from Eq. (\ref{eq_recurs_Ab}) with
$n_{\rm max}=70$.
The results agrees with numerical solutions for Eq. (\ref{eq_eff}) (see below)
when $\lambda\lesssim 0.01$. However, ignoring convection in the denominator of
Eq. (\ref{eq_eff}) due to the small $\lambda$ limit implies neglecting hydrodynamic
dissipation.
Therefore, the asymptotic efficiency in Eq. (\ref{eq_eff_asympt}) is
not strictly bounded by unity.
\subsection{Numerical results for the efficiency of swimming}
For a numerical evaluation of Eq. (\ref{eq_eff}) we truncate the expansion in
Legendre polynomials at $n_{\rm max} = 4$  as in Sec. \ref{sec_react_pol}.
Due to the linear response nature of our theory, numerator and denominator of
Eq. (\ref{eq_eff}) are both quadratic in $\Delta\mu_{\infty}$. Therefore, the equilibrium perturbation
driving the motion of the swimmer does not appear in the efficiency. However, the
results support the notion that the swimming efficiency increases away from
the quasi-equilibrium limit. 

\begin{figure}[ht]
\begin{center}
\includegraphics[scale=0.87]{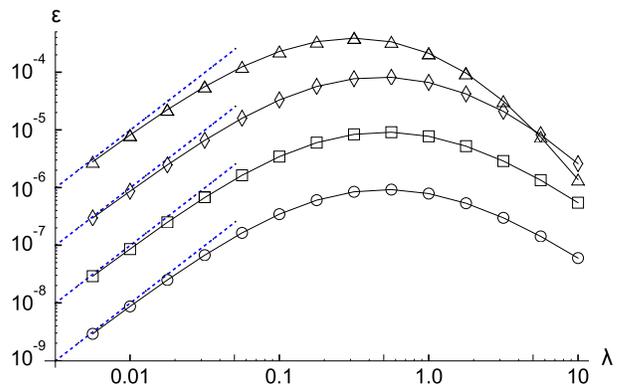}
\caption{\label{fig_4-1}The swimming efficiency $\epsilon$ as a function of $\lambda$
with constant $\Pe/\delta$ for $\Psi(y) = \exp\left[-(r-1)/\lambda \right]$ and $k_{\rm ab}=k_{\rm ba}=1$.
(Dashed lines) Asymptotic results calculated from Eq. (\ref{eq_eff_asympt}).
~($\triangle$) $\Pe/\delta=10$. ~($\lozenge$) $\Pe/\delta=1$.
~($\square$) $\Pe/\delta=0.1$. ~($\circ$) $\Pe/\delta=0.01$.}
\end{center} 
\end{figure}

\begin{figure}[ht]
\begin{center}
\includegraphics[scale=0.87]{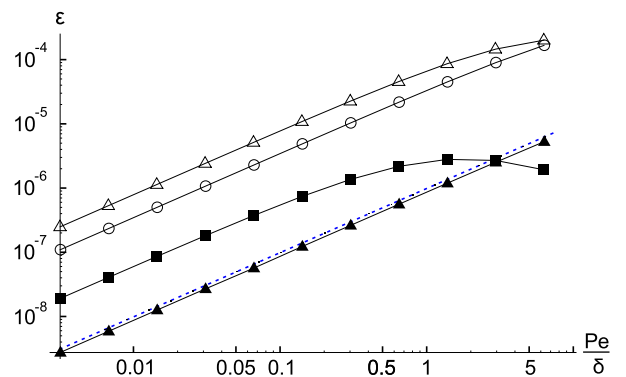}
\caption{\label{fig_4-2}The swimming efficiency $\epsilon$ as a function of
$\Pe/\delta$ with constant $\lambda$ for $\Psi(y) = \exp\left[-(r-1)/\lambda \right]$, $k_{\rm
ab}=k_{\rm ba}=1$.
(Dashed line) Asymptotic result for $\lambda=0.01$ calculated from Eq.
(\ref{eq_eff_asympt}).
~($\blacksquare$) $\lambda=10$. ~($\triangle$) $\lambda=1$.
~($\circ$) $\lambda=0.1$. ~($\blacktriangle$) $\lambda=0.01$.}
\end{center} 
\end{figure}

As seen in Figs. \ref{fig_4-1} and \ref{fig_4-2}, the swimming efficiency is 
proportional to $\lambda^2\,\Pe/\delta \sim L^4$ for $\lambda \ll 1$. This scaling is also evident from the prefactor 
in Eq. (\ref{eq_eff}). With a fixed Pecl\'et number, $\Pe/\delta >0$, the swimming efficiency decreases 
for $\lambda \gtrsim 1$ (Fig. \ref{fig_4-1}). Comparison with Fig. \ref{fig_2-1} shows that
the reduction in efficiency is due to the reduction of swimming speed in this regime. When, for $\lambda \gtrsim 1$, the swimming speed
does not increase $\sim\lambda^2$ , fixing the Pecl\'et number in Eq. (\ref{eq_eff}) 
introduces a scaling of $\epsilon$ with a negative power of $\lambda$. As an alternative, one could
remove the dependence of $\Pe$ on the interaction lengthscale $L$ by setting $\Pe/(\delta \lambda^2) = \mathrm{const}$
for Fig. \ref{fig_4-1}. This way of plotting the data would render the decrease of $\epsilon$ for $\lambda \gtrsim 1$ less pronounced. 

According to Eqns. (\ref{eq_Pe_def},\ref{eq_delta_active_def},\ref{eq_Uscale_def}), fixing
$\Pe/\delta$ and $\lambda$ implicitly sets an absolute equilibrium concentration scale. 
Therefore, Fig. \ref{fig_4-2} also suggests that the efficiency of
diffusiophoretic swimming increases with the absolute concentration scale for $\lambda \lesssim 1$. Only the asymptotic analytical
result for $\lambda = 0.01$ has been plotted in Fig. \ref{fig_4-2} because the curve for $\lambda = 0.1$ already showed significant deviations
from the numerical data.

\begin{figure}[ht]
\begin{center}
\includegraphics[scale=0.87]{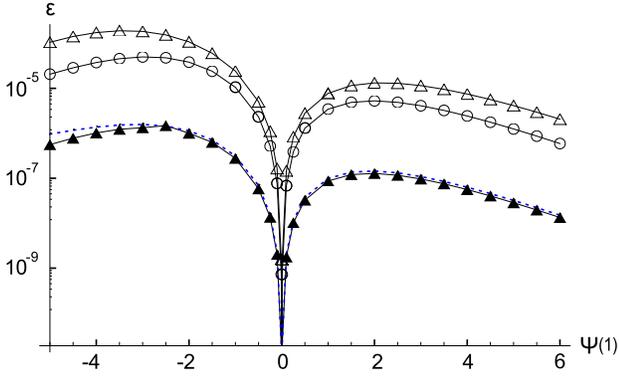}
\caption{\label{fig_4-3}Swimming efficiency $\epsilon$ vs. the strength of the 
interaction potential $\Psi(1)$, where $\Psi(y) = \Psi(1)\exp\left[-(r-1)/\lambda \right]$, $k_{\rm ab}=k_{\rm ba}=1$
and $\Pe/\delta=0.1$.
(Dashed line) Asymptotic result for $\lambda=0.01$ calculated from Eq.
(\ref{eq_eff_asympt}).
 ~($\triangle$) $\lambda=1$.
~($\circ$) $\lambda=0.1$. ~($\blacktriangle$) $\lambda=0.01$.}
\end{center} 
\end{figure}

Fig. \ref{fig_4-3} shows the dependence of the swimming efficiency on the strength of the interaction potential $\Psi(1)$. $\Psi(1)$ influences
both, the reaction rate and the swimming speed and therefore it has a nonlinear effect on $\epsilon$. Reaction induced concentration distortion plays an important role
for the shape of the curve, in particular for $\Psi(1)<0$ when $k_{+}$ can become much larger than unity. For $\Psi(1) \rightarrow 0$ the swimming efficiency vanishes.

\begin{figure}[ht]
\begin{center}
\includegraphics[scale=0.87]{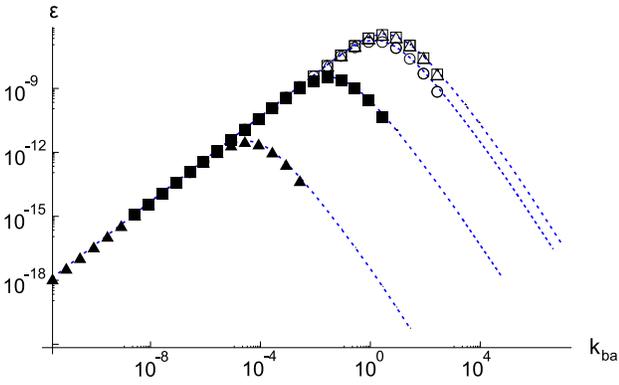}
\caption{\label{fig_4-4} Dependence of the swimming efficiency $\epsilon$ on the reaction rate constant $k_{\rm
ba}$ with fixed equilibrium constant
$K_{\rm a} \equiv \tc_{a}^{\rm eq}(\infty)/\tc_{b}^{\rm eq}(\infty)$ and $\Pe/\delta=0.1$,
$\lambda=0.01$, $\Psi(y) = \exp\left[-(r-1)/\lambda \right]$.
(Dashed lines) Asymptotic results calculated from Eq.
(\ref{eq_eff_asympt}).
~($\blacktriangle$) $K_{\rm a} = 10^{-5}$.
~($\blacksquare$) $K_{\rm a} = 10^{-2}$.
~($\circ$) $K_{\rm a} = 1$.
~($\square$) $K_{\rm a} = 10^{2}$.
~($\triangle$) $K_{\rm a} = 10^{5}$.
}
\end{center} 
\end{figure}

Figure \ref{fig_4-4} shows the swimming efficiency for fixed equilibrium constant $\tc_{a}^{\rm eq}(\infty)/\tc_{b}^{\rm eq}(\infty)$ and 
varying reaction rate $k_{\rm ba}$. For simplicity, we consider here only $\lambda =0.01 \ll 1$. 
Due to the truncation of the full numerical solution at Legendre polynomials of the order $n_{\rm max} = 4$ the error in the numerical data becomes large beyond the plotted range. 
The reaction induced concentration distortion again explains major features of the plotted curves. For $k_{\rm ba} \exp(-\Psi(1)) \ll 1$ and $k_{\rm ba} \exp(-\Psi(1))\lesssim \tc_{a}^{\rm eq}(\infty)/\tc_{b}^{\rm eq}(\infty)$ the 
reaction induced concentration distortion is negligible. Then $\epsilon$ rises
linearly with $k_{\rm ba}$ and the curves fall onto each other. For larger equilibrium constants with $\tc_{a}^{\rm eq}(\infty)/\tc_{b}^{\rm eq}(\infty) \gg 10$ the location of the maximum
of swimming efficiency becomes becomes independent of the equilibrium constant. On employing Eq. (\ref{eq_eff_asympt}) we here find the maximum of $\epsilon$ at $k_{\rm ba} \simeq 1.13 \exp\Psi(1)$.
 

\subsection{Efficiency of transport}
The definition of swimming efficiency in Eq. (\ref{eq_eff}) is based on the comparison
of energy dissipation per unit time. In certain practical applications of 
diffusiophoresis one might prefer other efficiency measures. It is, e.g., 
interesting to have a measure for the energetic cost of using active swimmers 
for transport between two locations. Then, it may be more appropriate to compare
energy 
dissipation per transport distance. For concreteness, we think here of a slab
geometry which the swimmers are to cross. Thereby, they move from one slab at 
$\tx=0$ to the other slab at $\tx=\tX >0$.
Since the swimmers are freely suspended, they will not swim straightly but
Brownian, translational and rotational, diffusion comes into play.
The translational diffusion has a constant $D_{\rm S} = kT/\left(6 \pi
\eta R\right)$ and the rotational diffusion happens on a timescale 
$\tau_{\rm rot} = 8 \pi \eta R^3/kT$. For long times $\ttt \gg \tau_{\rm rot}$, one can employ an effective
translational diffusion constant\cite{howse2007self,golestanian2009anomalous,ke2010motion,palacci2010sedimentation} where 
the effects of random diffusion and active, translational motion
are incorporated as 
$D_{\rm eff} = D_{\rm S}+\tU^2\,\tau_{\rm rot}/6$. This approach neglects
possible modifications of $D_{\rm S}$ and $\tau_{\rm rot}$ due to the active
processes. In our setup we expect the transport distance $\tX$ to be much larger than a
characteristic length of the random walk $\tX \gg D_{\rm eff}/|\tU|$. Employing 
reflecting boundary conditions at $\tx=0$,
the mean first passage time $\ttt_{\rm mf}$ to reach $\tx = \tX$ is simply given
by
$\ttt_{\rm mf} =
\tX^2/\left(2D_{\rm eff}\right)$. The energy consumed by the swimmer during the
time
$\ttt_{\rm mf}$ can be estimated by $\tP_{\rm in}\,\ttt_{\rm
mf}$. 
For an energetic comparison, one might consider dragging a passive particle
directly across $\tX$. The corresponding dissipated work would be $6 \pi\eta R 
|\tU|\,\tX$. Hence, we define a new efficiency of transport as
\begin{equation}
 \begin{split}
 \epsilon_{\rm transp} \equiv &\frac{6 \pi \eta R |\tU| \, \tX}{\tP_{\rm in}\,
\ttt_{\rm
mf}} =
\epsilon\,\frac{\tX}{|\tU|\, \ttt_{\rm mf}},
 \end{split}
\end{equation}
which we relate on the right hand side to the efficiency of swimming $\epsilon$ through Eq. (\ref{eq_eff}).
On employing the result for the mean first passage time $\ttt_{\rm mf}$ given in the text above we find
\begin{equation}
 \epsilon_{\rm transp}=\frac{\epsilon}{\tX}\,\frac{2 D_{\rm eff}}{|\tU|}=\frac{\epsilon}{\tX} \left(
\frac{2 D_{\rm S}}{|\tU|} + \frac{|\tU| \tau_{\rm
rot}}{3}\right).
\end{equation}
This dependence $\sim 1/\tX$ of the relative energetic cost of direct transport 
compared to enhanced diffusion is a quite generic result. It emphasizes
the necessity to impose a directionality on the active motion if the transport
distance is to be of macroscopic size. 
\section{Discussion and conclusion}
In the present publication we take a close look at the dynamics and efficiency
of diffusiophoretic swimming in the linear response regime. 
In Sec. \ref{sec_swimming_speed} we analyze the role of the surface
interaction potential with regard to the dynamics of the swimmer. 
If the lengthscale of the interaction potential is not negligible compared to
the swimmer radius ($\lambda \gtrsim 0.01$), 
the details of the interaction become important. Analytical corrections in
powers of $\lambda$ to the lowest order swimming speed include various moments
of the solute concentration and also the effect of solute convection. These
corrections occur for active diffusiophoresis in a  qualitatively similar way as
for passive diffusiophoresis\cite{anderson1984diffusiophoresis}. A particular
feature of active diffusiophoresis, however, is that the emission of solutes can
cause an increase of speed for repulsive surface interactions. The enhanced speed 
reflects the effect of ''producing'' solutes directily inside the region from which they are repelled. 

In Sec. \ref{sec_react_pol} we employ a linear reaction scheme to model the
active conversion of solutes at the swimmer's surface. Concentration
dependent boundary conditions result in a strong modification of the
concentration fields as compared to fixing the flux of solutes at the boundary. 
This reaction induced concentration distortion also occurs
in the limit $\lambda \ll 1$ and has a pronounced effect on the swimming
speed. In the linear response regime it causes a reduction of the swimming
speed, independent of the absolute concentration levels. If the reaction rate becomes
nonlinear in the concentrations, a measurable, concentration dependent, speed
modification is to be expected. The whole effect may vanish only in the strongly
reaction limited regime of very high substrate concentrations. 

In Sec. \ref{sec_energetics} we present results concerning the efficiency of
diffusiophoretic swimming. The active motion considered in the present
publication is energetically
quite distinct from classical phoresis in external gradients
\cite{anderson1989colloid} .
For the latter systems, the absolute level of concentration changes when the
particle moves in an externally applied gradient. In our case the absolute
concentrations remain, 
on average, the same; which implies, that our system can be held in a true steady
state. The studied
model includes three modes of entropy production namely hydrodynamic
dissipation, dissipation in the chemical reaction and
entropy production through diffusion of solutes. While the former two can be
recognized immediately as contributions to the power input, the latter becomes
only meaningful when considering an external apparatus, which converts free
energy exchange at the outer boundaries of the system into work. Identification
of the power consumption with the free energy exchange at the outer boundaries of the system
is a 
consequence of the spatial modelling of the process. It is a natural extension
to the commonly employed formalism for
biomolecular motors\cite{parmeggiani1999energy}, where the local free energy
exchange rate is taken as work
input.
It is pivotal for the energy balance to recognize of the importance of
convection.
For illustration, we think of a situation where the reaction rates would be fully
independent of the fluid motion. This would mean that the net exchange of
solutes 
with the external apparatus would
be independent of the motion of the particle. Thus, according to
Eq.(\ref{eq_pin_model}), 
the power input would be independent of the hydrodynamic dissipation, which
violates energy conservation. Within this model, the only way how the particle
speed may feed-back on the chemical reaction is via a convective modification of the
concentrations. This link leads, together with the mass-action law, to a proper energy balance.

Relying on the robustness of the scaling of the efficiency 
$\epsilon \sim D_s/D \times L^4 \hc/R$ (Eq. (\ref{eq_eff_asympt})), one 
can employ experimental values for catalytic swimmers to estimate their
efficiency. 
For swimmers with size in the $1 \, \mu m$ range, we have $D_s/D \simeq
10^{-4}$. Given a concentration
scale\cite{howse2007self} of $\hc \simeq 10^7\, 1/\mu m^3$ and an 
interaction length of $L\simeq 1\,\rm nm$ we have $L^4 \hc/R = 10^{-5}$. 
Together we find $\epsilon \simeq 10^{-9}$. This number corresponds 
to the experimental estimate for the swimmers of Paxton et.
al\cite{paxton2005motility}. The agreement may, however, be incidental since these 
swimmers are operating mainly with a self-electrophoretic mechanism instead of the 
diffusiophoretic mechanism studied
here. Naturally, if the concentration perturbation $\hc$ is caused by a large 
chemical potential difference, this  may also affect the efficiency 
as seen from Eq. (\ref{eq_eff_asympt}). 

In conclusion, we expect the main relevance of this work to lie in the scaling
predictions and in the identification of the main trends in the numerical data. A more detailed
modeling of specific phoretic mechanisms is challenging, but desirable. One
interesting question is to replace the conservative potential $\Psi$ by more
realistic, angular dependent surface interactions. In reality, the
solute-swimmer
interactions may even transfer energy between the fluid and microscopic degrees
of freedom on the swimmer's surface. Thus, they can acquire a ''dissipative
nature'', possibly resulting in an effective modification of the viscosity
around the swimmer. Also, lateral 
diffusion of solutes adsorbed to the surface of the swimmer could be taken into 
account here. Issues like hydrodynamic surface slip \cite{ajdari2006giant} may further
complicate the situation. For a full description of the behavior of
catalytically driven swimmers one must also go beyond the linear response theory. 
This refinement of the model will particularly
effect the reaction induced concentration distortion. Finally, the 
predictions for the scaling of the swimming speed and efficiency offer the 
possibility of being experimentally tested, which could add
interesting new facets to the picture of diffusiophoresis presented here.
\appendix
\section{Simple model for the dilute solution}\label{sec_dilute_solution_Appendix}
In this Appendix we provide a simple, but detailed derivation of the model fluid
employed 
in the main part of the text. We denote the molecular masses of the solutes and
solvent by $m_a$,$m_b$ and
$m_s$. The molecular volumes are constants, denoted by $v_a$,$v_b$ and $v_s$
respectively. The volume variable is written as $\tilde{V}$. We define a local, 
position dependent free energy per unit volume $\tf\equiv \tilde{F}/\tilde{V}$
as
\begin{equation}
\begin{split}
&\tf(\tc_a,\tc_b,\tc_s) \equiv \phi \tPsi_0 + \tc_b v_b \tPsi_1 + \tc_b
\epsilon_b+ \tc_a \epsilon_a +
\frac{\chi}{2}\left(\phi -1\right)^2 + \\
&kT\left[\tc_a \log\left(\frac{\tc_a v_a}{\phi}\right)+\tc_b
\log\left(\frac{\tc_b v_b}{\phi}\right) +\tc_s\log\left(\frac{\tc_s
v_s}{\phi}\right) \right],
\end{split}
\end{equation}
where $\phi\equiv\tc_a v_a + \tc_b v_b +\tc_s v_s$.
Here the surface interactions between the fluid constituents and the swimmer are
incorporated through the potential energies per volume $\tPsi_0$ and $\tPsi_1$.
$\epsilon_{a}$,$\epsilon_{b}$
are the internal energies of the solutes of type $a$ and $b$ respectively. 
The local, hydrostatic pressure is calculated as 
\begin{equation}
\begin{split}
 \tilde{p} = -\frac{\rmd \left(\tf\,\tilde{V}\right)}{\rmd \tilde{V}} &= -\tf
+\frac{\p \tf}{\p \tc_a} \tc_a +\frac{\p \tf}{\p \tc_s} \tc_s+\frac{\p \tf}{\p
\tc_s} \tc_s = \\
&= \frac{\chi}{2}\left[\phi^2-1\right].
\end{split}
\end{equation}
In the incompressible limit one has $\phi=\left(\tc_a v_a + \tc_b v_b +\tc_s
v_s\right)
\approx 1$ and the pressure can be expanded as $\tilde{p} \approx
\chi\left[\phi-1\right]$. The generalized
chemical
potentials $\tmu_i'\equiv \tmu_i + v_i\left(\tPsi_0+\delta_{i,b}\tPsi_1\right) =
\p \tf/\p \tc_i $ become, with $\tc_{ all} \equiv \left(\tc_a +\tc_b +\tc_s\right)$,
\begin{align}
 \begin{split}
  \tmu'_s \approx & kT \left[\log\left(\tc_s v_s \right) 
-\tc_{ all} v_s \right] + \\
	&v_s \left(\tPsi_0 +\tp\right) +kT \left( v_s \tc_{ all}- 1 \right)\left(\phi-1\right),
 \end{split}\\
\begin{split}
\tmu'_a \approx & kT \left[\log\left(\tc_a v_a \right)  -\tc_{ all}
v_a\right] + \epsilon_a+ \\
&v_a\left(\tPsi_0 +\tp\right) +kT \left( v_a \tc_{ all}- 1 \right)\left(\phi-1\right),
\end{split}\\
\begin{split}
 \tmu'_b \approx & kT \left[\log\left(\tc_b v_b \right) -\tc_{ all}
v_b\right] + \epsilon_b + \\
&v_b\left(\tPsi_0 + \tPsi_1 +\tp\right) +kT \left( v_b \tc_{ all}- 1 \right)\left(\phi-1\right).
\end{split}
\end{align}
We now apply the highly simplifying assumption 
of symmetric solute constituents
\begin{align}
  &v_a \simeq v_b \simeq  v_s, \label{eq_all_volumes_same}\\
  &m_a \simeq m_b \simeq  m_s \label{eq_app_same_spec_vol}
\end{align}
through which particle and mass fluxes become equivalent. Also, we assume that the volume fraction of 
solvent is much larger than that of the solute 
\begin{equation}
 \tc_a v_a +\tc_b v_b \ll 1. \label{eq_dilute_limit}
\end{equation}
The linear phenomenological equations, 
linking a diffusive flux $\tbj$ to gradients in chemical potential, read
\begin{align}
\begin{split}
\tbj_b &= L_{bb} \nabla \left(\frac{\tmu'_b}{m_b} -\frac{\tmu'_s}{m_s}\right) +
L_{ba} \nabla
\left(\frac{\tmu'_a}{m_a} -\frac{\tmu'_s}{m_s}\right),\\
\tbj_a &= L_{aa} \nabla \left(\frac{\tmu'_a}{m_a} -\frac{\tmu'_s}{m_s}\right) +
L_{ab} \nabla
\left(\frac{\tmu'_b}{m_b} -\frac{\tmu'_s}{m_s}\right). \label{eq_phen_relations}
\end{split}
\end{align}
The flux of the solvent can then be calculated from the condition
\begin{equation}
\sum_{i=a,b,s} m_i \tbj_i = 0.
\end{equation}
The coefficients $L_{m n}$ in Eq. (\ref{eq_phen_relations}) relate gradients in chemical
potential to dissipation. For $\tc_a v_a +\tc_b v_b \ll 1$, the diagonal 
coefficients $L_{a a}$ and $L_{b b}$ are proportional to the product of a solute volume fraction
and the solvent volume fraction. The solvent volume fraction is close to unity and
we set $L_{bb} =-m_b \tc_b D/kT $ and $ L_{aa} = -m_a \tc_a D/kT $. 
The cross-coefficents, on the other hand, are proportional to the product of the two solute 
volume fractions. They obey the Onsager relation $L_{ab} = L_{ba}$. To lowest order in solute volume fractions 
we therefore have $L_{ab} = L_{ba} \approx 0$ when Eq. (\ref{eq_dilute_limit}) holds. 
On employing Eqns. (\ref{eq_all_volumes_same}-\ref{eq_phen_relations})
we finally obtain
\begin{align}
\tbj_a &= -D \nabla \tc_a,\\
\tbj_b &= -D \left(\nabla \tc_b + \frac{\tc_b}{kT}\nabla\tPsi\right), 
\end{align}
where the surface potential acting on each molecule of type $b$ is defined as
$\tPsi \equiv v_b \tPsi_1$. Due to the assumption of similar specific volumes of the molecules in 
Eqns. (\ref{eq_all_volumes_same},\ref{eq_app_same_spec_vol}) the pressure becomes a purely hydrodynamic quantity. 
We have in steady state the common incompressibility condition $\nabla\cdot\tbv=0$ since
\begin{align}
\begin{split}
 0=&\nabla\cdot\left[\left(m_a \tc_a+m_b \tc_b+m_s \tc_s\right)\tbv\right]
\approx \\
 &\frac{m_s}{v_s}\nabla\cdot\left[\left(\tc_a v_a +\tc_b v_b +\tc_s
v_s\right)\tbv \right] \approx \frac{m_s}{v_s}\nabla\cdot\tbv.
\end{split}
\end{align}
\section{Diffusiophoretic swimming for $\lambda \ll 1$
}\label{sec_app_asympt_small_L}
In this Appendix the speed $U_0$ of active swimmers is calculated analoguous to
work by Anderson and Prieve \cite{anderson1984diffusiophoresis} for passive
swimmers. The diffusion equation (\ref{eq_c_b_dgl},\ref{eq_convect_lin_response}) becomes in spherical coordinates 
\begin{equation}
\begin{split}
&\p_r^2 c(r,\theta) + \left(\Psi'(r) +\frac{2}{r}\right)\p_r c(r,\theta) \\
&+\left(\Psi''(r) +\frac{2}{r}\Psi'(r) \right)c(r,\theta) + \frac{1}{r^2 \sin \theta} \p_{\theta}\left(\sin\theta\,\p_{\theta}c(r,\theta)\right)\\
&-\frac{\Pe}{\delta} \hbe_r\bv\, \p_r\, e^{-\Psi(r)}   =0. \label{eq_app_diff}
\end{split}
\end{equation}
Due to the radial symmetry of $\Psi(r)$, only the dipole moment of 
the concentration field can move the particle.  In absence of a coupling between
different spherical harmonics it is therefore sufficient to consider the field 
contributions  which are $\sim \cos\theta$. 
We split the concentration field of type $b$ solute into 
an inner field $C^{i}(y) \cos\theta$ and an outer field $C^{o}(r) \cos\theta$.
$C^{i}(y)$ is written in terms of the inner variable $y\equiv
\left(r-1\right)/\lambda$. The outer field lies in the region with $r \gg
\lambda + 1$ where the effect of the surface potential $\Psi$ is negligible. Eq.
(\ref{eq_app_diff}) thus becomes here
in the outer region
\begin{equation}
\p_r^2 C^{o}(r) + \frac{2}{r}\p_r C^{o}(r) - \frac{2}{r^2} C^{o}(r) = 0. \label{eq_laplacian_outer}
\end{equation}
For active diffusiophoresis, the concentration perturbation must vanish 
far away of the swimmer, Therefore, the solution of Eq. (\ref{eq_laplacian_outer}) is $C^o(r) = A/r^2$, where the coefficient $A$
needs to be determined by matching with the far field behavior of
$C^i$. The smallness of $\lambda$ suggests an expansion of the inner and outer
fields as 
\begin{align}
&C^i = C^i_0(y) +C^i_1(y) \lambda + C^i_2(y) \lambda^2 + O(\lambda^3),\\
\begin{split}
&C^o =\frac{1}{r^2}\left(A_0 + A_1 \lambda + A_2 \lambda^2 + \ldots\right)=
A_0 + \\ 
&\left(A_1 - 2 A_0 y \right) \lambda + \left( A_2 - 2 A_1 y + 3 A_0 y^2\right)
\lambda^2 + O(\lambda^3).
\end{split}
\end{align}
In the inner region we substitute $r$ by $\lambda y + 1$ in Eq.
(\ref{eq_app_diff}) and expand for small $\lambda$. 
The resulting equations for the two lowest coefficients $C^i_0$ and $C^i_1$ are 
\begin{align}
&\p_y\left(\p_y C^i_0(y)  + C^i_0(y) \p_y\Psi\right) = 0, \\
&\p_y\left(\p_y C^i_1(y)  + C^i_1(y) \p_y\Psi\right) + 2 \left(\p_y C^i_0(y)  +
C^i_0(y) \p_y\Psi\right)= 0.
\end{align}
The boundary condition Eqns. (\ref{eq_j_bc},
\ref{eq_alpha_const_bc}) yield for the coefficients of $C^i(y)$
\begin{align}
&\frac{- \lambda}{\lambda} \left(\p_y C_1^i(y) + C^i_1(y)  \p_y
\Psi\right)|_{y=0} = \frac{\kappa R}{D \hc}=1;\\
&\frac{-\lambda^n}{\lambda} \left(\p_y C^i_n(y)  + C^i_n(y)
\p_y\Psi\right)|_{y=0} = 0 & & \text{when} & &n \neq 1.
\end{align}
Employing the above equations, the concentration is 
\begin{equation}
\begin{split}
 C^i = &e^{-\Psi} a_0 + \lambda\, e^{-\Psi} \left[a_1 -  \int_0^y\left(
e^{\Psi(y')} -1 \right)\rmd y' -  y\right] +\\
 &O\left(\lambda^2\right).
\label{eq_app_ci1}
\end{split}
\end{equation}
The $O\left(\lambda\right)$ term diverges for $y\rightarrow \infty$. In order to
make this divergence explicit, we have removed it from the integral in Eq.
(\ref{eq_app_ci1}) by subtracting $1$. The unknown constants in the inner and 
outer solution are determined by matching them asymptotically\cite{van1975perturbation} through
$C^o_n\left(y\rightarrow 0\right) = C^i_n\left(y\rightarrow \infty\right)$. The resulting
conditions, which must be valid for all $y$, are
\begin{align}
A_0 & = a_0,\\
A_1 - 2 A_0 y & = a_1 -  \int_0^{\infty}\left(
e^{\Psi(y')} -1 \right)\rmd y' -  y. 
\end{align}
This yields the lowest order coefficients $a_0 = A_0 = 1/2$ and the
innermost concentration field is thus given by $C^i \approx 
\exp(-\Psi)/2$. 

In order to calculate the fluid flow near the surface of the
swimmer, we employ the Stokes equation (\ref{eq_stok_dgl}) and assume that the 
body force vanishes in the outer region. Applying the curl to Eq.
(\ref{eq_stok_dgl}) and defining a stream function $S(r)$ via $\bv = \nabla
\times \left(\sin\theta S(r)/r\,\mathbf{\hat{e}}_{\varphi}\right)$ the Stokes
equation becomes
\begin{align}
\p^4_rS(r) - \frac{4\,\p^2_r S(r)}{r^2}+ \frac{8 \,\p_r S(r)}{r^3}- \frac{8
S(r)}{r^4} = -\frac{C^i}{\lambda^2}\, \p_r \Psi(r) \label{eq_app_str_fkt_dgl}.
\end{align}
In the outer region, where the body force vanishes, we have the stream function
$S^o =X/r+Y\,r+Z\,r^2$. The constants $X$,$Y$ and $Z$ are expressed as a power
series of $\lambda$. The fluid velocity in the outer region thus is
\begin{align}
\frac{\hat{\mathbf{e}}_{\theta}\bv^o}{\sin\theta} &= \frac{-\p_r S^o(r)}{r} = - 2 Z_0 - Y_0 + X_0 +
O\left(\lambda\right),\\
\frac{\hat{\mathbf{e}}_{r}\bv^o}{\cos\theta} &= \frac{2 S^o(r)}{r^2} = 2 Z_0 + 2 Y_0 + 2 X_0 +
O\left(\lambda\right).
\end{align}
In the inner region, we expand Eq. (\ref{eq_app_str_fkt_dgl}) for small $\lambda$ and 
insert $C^i(y)$ for the concentration field. The leading order
differential equation for the stream function near the surface of the swimmer $S^i$ reads
\begin{align}
\frac{1}{\lambda} \p_y^4 S^i(y) = -\frac{e^{-\Psi}}{2} \p_y \Psi(y).
\end{align}
This yields
\begin{align}
\begin{split}
 S^i(y) = & k_0 + l_0\, y + m_0\, y^2 + n_0\, y^3 - \frac{\lambda }{2}
h(y) \\
 &\lambda\left(k_1 + l_1\, y + m_1\, y^2 + n_1\, y^3\right)
+O\left(\lambda^2\right);
\end{split}\\
h(y) \equiv & \int_0^y\int_0^{y'}\int_{y''}^{\infty}
\left(e^{-\Psi(y''')}-1\right)\rmd y''' \rmd y'' \rmd y'.
\end{align}
The fluid flow in the inner region becomes
\begin{align}
\begin{split}
 &\frac{\hat{\mathbf{e}}_{\theta}\bv^i}{\sin\theta} \approx   -\frac{1}{\lambda}\left(l_0\,  + 2
m_0\,y + 3 n_0\, y^2\right) + \frac{1}{2}  \p_y h(y) +\\
  &\left( -l_1 + l_0 y - 2 m_1 y + 2 m_0 y^2 - 3 n_1 y^2 + 3 n_0 y^3\right)  +O\left(\lambda\right),
\end{split}\\
&\frac{\hat{\mathbf{e}}_{r}\bv^i}{\cos\theta}  \approx   2\left(k_0 + l_0\, y + m_0\, y^2 + n_0\, y^3\right)+O\left(\lambda\right).
\end{align}
Due to the no slip boundary conditions on the surface of the swimmer we have
$k_0 =l_0=l_1= 0$. The far field boundary condition on the outer velocity field
$\bv|_{r \rightarrow \infty} = -U \hat{\mathbf{e}}_{z}$ yields $Z=-U/2$.
Matching the lowest order velocities through $\bv^i_n\left(y\rightarrow \infty\right)=\bv^o_n\left(y\rightarrow 0\right)$, as done for the concentrations above, we
find
\begin{align}
0&=-\frac{1}{\lambda}\left(2 m_0\, y + 3 n_0\, y^2\right), \label{eq_m1}\\
- 2 Z_0 - Y_0 +X_0 &= \frac{K_1}{2} -2 m_1 y +2 m_0 y^2 -3 n_1 y^2 + 3 n_0 y^3  ,\label{eq_m2}\\
2 Z_0 + 2 Y_0 + 2 X_0 &= 2m_0\, y^2 + 2n_0\, y^3, \label{eq_m3}
\end{align}
where we have used $\p_y h(y)|_{y\rightarrow
\infty} = K_1$ (see Eq. (\ref{eq_Kn_def})). From Eq. (\ref{eq_m1}) we
deduce that $m_0=n_0 = 0$. On imposing
the physical constraint that $\p_y h(y)$ remains finite for $y\rightarrow \infty$
we conclude that $m_1 = n_1 = 0$ in order to 
avoid divergence of the right hand side of Eq. (\ref{eq_m2}). 
Finally, Eqns. (\ref{eq_m2},\ref{eq_m3}) yield together the two last constants
$X_0=\left( K_1 - U_0\right)/4$ and $Y_0=\left( 3
U_0-K_1\right)/4$. This fully determines the lowest order velocity fields. To
further relate the speed of the swimmer $U_0$ to the balance of forces, one only
needs to consider the Stokeslet $\bv^o \sim 1/r$, whose long range nature
reflects the presence of external forces. For a free swimmer the Stokeslet
vanishes and we therefore have $Y_0=0$. This condition determines the lowest order
swimming speed
\begin{equation}
U_0 =  \frac{K_1}{3}.
\end{equation}
The first correction of $O(\lambda^3)$ in a series expansion of $\tU$ (see Eq. (\ref{eq_active_U_smol})) is
an extension of the scheme presented here. 
\section{Energy balance in linear, nonequilibrium thermodynamics}\label{sec_app_energy_balance}
This Appendix substantiates the definition of the power input 
in Sec. \ref{sec_energetics} by providing more details about the underlying assumptions and calculations.
As explained in Sec. \ref{sub_sec_energy_balance}, we assume that an   
external apparatus is connected to the system. The apparatus consumes energy and keeps the system
in steady state. The ensemble of system and apparatus 
does not exchange matter but only work and heat with the external world.
The first law of thermodynamics for the ensemble reads
\begin{equation}
 \tP_{\rm in} = \tP_{\text{out}}+\delta \tQ
\end{equation}
where $\tP_{\text{out}}$ is the work output of the system. We set $\tP_{\text{out}}=0$ 
because no external force acts on the swimmer. Within the framework of classical, linear, nonequilibrium thermodynamics
\cite{degroot1984, julicher2009generic}, the heat flow $\delta \tQ$ balances in steady state the entropy production 
inside the system and the apparatus. We can employ a local definition of the entropy production rate to rewrite the first law
\begin{equation}
\begin{split}
 \tP_{\rm in} =\delta \tQ =  \int T \left(
\tTheta_{\rm sys} +\tTheta_{\rm app} \right)
\,\rmd \tV \label{eq_pin_general}
\end{split}
\end{equation}
with $\tTheta$ being the entropy production rates per unit volume. Since the external apparatus is ideal, we have $\tTheta_{\rm app}=0$. 
For the entropy production of the system we employ an established formula\cite{degroot1984}, which derives from a local equilibrium assumption,
\begin{equation}
\begin{split}
\tTheta_{\rm sys} =& \frac{1}{T} 2\eta
\tE:(\nabla \tbv) - \frac{1}{T} \sum_i \tbj_i \nabla \left( \tmu_i +  \tPsi_i \right) \\
&-\frac{1}{T} \sum_{k i} \left( \tmu_i +\tPsi_i \right) \talpha_{i k},
\label{eq_entropy_prod}
\end{split}
\end{equation}
where $\talpha_{i k}$ is the rate at which the reaction with index $k$ changes the concentration 
of species $i$. For the model system considered in Sec. \ref{sec_energetics}, the entropy production
rate is rewritten by employing the Stokes equation Eq. (\ref{eq_stok_dim}) and the chemical reactions defined in Eqns.
(\ref{eq_bc_mass_action_b},\ref{eq_bc_mass_action_a})
\begin{equation}
 \begin{split}
  T \tTheta_{\rm sys} \,\times \frac{R^2}{kT \hc D} =& -\left(\bj_b+\Pe \bv \frac{\tc_b}{\hc}\right)\nabla\Psi  - \sum_i \bj_i \nabla  \mu_i\\
  &+ \delta \left( r-1 \right) g(\theta) \left(\mu_a-\Psi -\mu_b\right)
\label{eq_phoretic_theta}.
 \end{split}
\end{equation}
In order to calculate $\tP_{\rm in}$, we assume that the chemical potentials $\tmu_a$,$\tmu_b$
and $\tmu_s$ are fixed by the external apparatus at $\tr \rightarrow \infty$. 
This, together with Eq. (\ref{eq_all_volumes_same}), 
implies that the concentrations at the outer boundaries of the system are fixed.
To facilitate reference, we note here the steady state diffusion equation 
\begin{equation}
 \nabla\cdot\tbJ_i=\nabla\cdot \left(\tbj_i+ \tbv \tc_i\right) = 0
\label{eq_ref_diff_dgl}.
\end{equation}
The solute flux boundary conditions at $\tr = R$ where $\tbv =0$ are given by
\begin{align}
\begin{split} 
\hbe_r \tbj_b|_{\tr=R}&= \frac{D \hc}{R}\, g(\theta),\\
\hbe_r \tbj_a|_{\tr=R} &=-\hbe_r \tbj_b|_{\tr=R},\\
\hbe_r \tbj_s|_{\tr=R} &=0. \label{eq_ref_bcs}
\end{split}
\end{align}
Inserting the entropy production rate Eq. (\ref{eq_phoretic_theta}) into the
expression for the power input 
Eq. (\ref{eq_pin_general}) and using Eq. (\ref{eq_ref_diff_dgl}) for the solute
of type $b$ we find
\begin{equation}
 \begin{split}
   \int T \tTheta_{\rm sys}  \rmd \tV =   -\int[  
\nabla\cdot\left[\left(\tbj_b+\tbv \tc_b\right)  \tPsi\right] + \sum_i \tbj_i
\nabla  \tmu_i ]\,\rmd \tV \\
+ \int \frac{D \hc }{R}  g(\theta) (\tmu_a-\tPsi -\tmu_b) \,\rmd
\tA_{\tr=R}. 
 \end{split}
\end{equation}
The potential $\tPsi$ now only occurs in surface integrals. $\tPsi(\tr)$ decays
quickly for $\tr \rightarrow \infty$. Furthermore, using Eq. (\ref{eq_ref_bcs})
 leads to a cancelation of surface integrals at $\tr=R$ containing $\tPsi$. We
are thus left with
\begin{equation}
 \begin{split}
   \int T \tTheta_{\rm sys}  \rmd \tV =   -\int [ \tbj_b \nabla  \tmu_b+  \tbj_a \nabla 
\tmu_a +\tbj_s \nabla  \tmu_s]\,\rmd \tV \\
+ \int \frac{D \hc }{R}  g(\theta) \left(\tmu_a-\tmu_b\right)\, \rmd \tA_{\tr=R}
\label{eq_ent_zw_erg}.
 \end{split}
\end{equation}
The Gibbs-Duhem equation, which is consistent with our microscopic model, reads
$0 =\sum_i \tc_i\nabla\tmu_i-\nabla \tp$. Multiplication
of this equation by the center of mass flow $\tbv$ and insertion into
Eq.(\ref{eq_ent_zw_erg}) yields 
\begin{equation}
 \begin{split}
   \int T \tTheta_{\rm sys}  \rmd \tV =   &-\int  [\sum_i\left(\tbj_i +\tbv
\tc_i\right)\nabla\tmu_i-\tbv \nabla \tp]\, \rmd \tV \\
&+ \int \frac{D \hc }{R}  g(\theta) \left(\tmu_a-\tmu_b\right) \,\rmd \tA_{\tr=R}
\\
=&-\int [ \sum_i\tbJ_i\tmu_i-\tbv \tp ]\, \hbe_r \rmd \tA_{\tr=\infty} \\
&+\int  [\sum_i \,\hbe_r \tbj_i\tmu_i +  \frac{D \hc }{R}  g(\theta)
\left(\tmu_a-\tmu_b\right)]\, \rmd \tA_{\tr=R} \label{eq_ent_zw_erg2},
 \end{split}
\end{equation}
where the diffusion equations and $\nabla\cdot\tbv=0$ were used to produce the
boundary integrals. Due to the boundary conditions Eq. (\ref{eq_ref_bcs}), the
last integral
in Eq. (\ref{eq_ent_zw_erg2}) vanishes. Since the swimmer is not subjected to
external forces, the fluid flow contains no Stokeslet. With the assumptions in Eqns.
(\ref{eq_all_volumes_same}, \ref{eq_app_same_spec_vol}) the
pressure does also not depend on local concentrations. Therefore, the boundary
work of the pressure vanishes $\int \tbv \tp \,\rmd \tbA_{\tr=\infty}
\rightarrow 0$ and can consequently be dropped in Eq. (\ref{eq_ent_zw_erg2}). 
For our model, we conclude that it does not make a difference whether we consider
a pressurized system (Gibbs free energy) or a system with fixed volume (free
energy). The power input becomes
\begin{equation}
 \begin{split}
   \int T \tTheta_{\rm sys}  \rmd \tV =&-\int  \sum_i\tbJ_i\tmu_i \,\hbe_r \rmd
\tA_{\tr=\infty}. \label{eq_ent_zw_erg3}
 \end{split}
\end{equation}
Employing the chemical potentials from Appendix
\ref{sec_dilute_solution_Appendix} with the assumption of diluteness ($v_a \tc_a
+v_b \tc_b \ll1$) and Eq. (\ref{eq_all_volumes_same}) 
along with Eqns. (\ref{eq_ref_diff_dgl},\ref{eq_ref_bcs}) we find 
\begin{align}
\begin{split}
&-\int  \sum_{i=a,b,s}\tbJ_i\tmu_i \,\hbe_r\rmd \tA_{\tr=\infty} = -\int 
\sum_{i=a,b}\tbJ_i\tmu_i \,\hbe_r\rmd \tA_{\tr=\infty}=\\
&\int \tbJ_b\left(kT\log\left[\tc_a/\tc_b\right] +
\epsilon_a-\epsilon_b\right)\,\hbe_r \rmd \tA_{\tr=\infty}.
\end{split}
\end{align}
This formula is employed in linearized form for Eq. (\ref{eq_pin_model}).

\begin{thebibliography}{10}%
\makeatletter
\providecommand \@ifxundefined [1]{%
 \ifx #1\undefined \expandafter \@firstoftwo
 \else \expandafter \@secondoftwo
\fi
}%
\providecommand \@ifnum [1]{%
 \ifnum #1\expandafter \@firstoftwo
 \else \expandafter \@secondoftwo
\fi
}%
\providecommand \enquote [1]{``#1''}%
\providecommand \bibnamefont  [1]{#1}%
\providecommand \bibfnamefont [1]{#1}%
\providecommand \citenamefont [1]{#1}%
\providecommand\href[0]{\@sanitize\@href}%
\providecommand\@href[1]{\endgroup\@@startlink{#1}\endgroup\@@href}%
\providecommand\@@href[1]{#1\@@endlink}%
\providecommand \@sanitize [0]{\begingroup\catcode`\&12\catcode`\#12\relax}%
\@ifxundefined \pdfoutput {\@firstoftwo}{%
 \@ifnum{\z@=\pdfoutput}{\@firstoftwo}{\@secondoftwo}%
}{%
 \providecommand\@@startlink[1]{\leavevmode}%
 \providecommand\@@endlink[0]{}%
}{%
 \providecommand\@@startlink[1]{%
  \leavevmode
  \pdfstartlink
   attr{/Border[0 0 1 ]/H/I/C[0 1 1]}%
   user{/Subtype/Link/A<</Type/Action/S/URI/URI(#1)>>}%
  \relax
 }%
 \providecommand\@@endlink[0]{\pdfendlink}%
}%
\providecommand \url  [0]{\begingroup\@sanitize \@url }%
\providecommand \@url [1]{\endgroup\@href {#1}{\urlprefix}}%
\providecommand \urlprefix [0]{URL }%
\providecommand \Eprint[0]{\href }%
\@ifxundefined \urlstyle {%
  \providecommand \doi [1]{doi:\discretionary{}{}{}#1}%
}{%
  \providecommand \doi [0]{doi:\discretionary{}{}{}\begingroup
  \urlstyle{rm}\Url }%
}%
\providecommand \doibase [0]{http://dx.doi.org/}%
\providecommand \Doi[1]{\href{\doibase#1}}%
\providecommand \bibAnnote [3]{%
  \BibitemShut{#1}%
  \begin{quotation}\noindent
    \textsc{Key:}\ #2\\\textsc{Annotation:}\ #3%
  \end{quotation}%
}%
\providecommand \bibAnnoteFile [2]{%
  \IfFileExists{#2}{\bibAnnote {#1} {#2} {\input{#2}}}{}%
}%
\providecommand \typeout [0]{\immediate \write \m@ne }%
\providecommand \selectlanguage [0]{\@gobble}%
\providecommand \bibinfo [0]{\@secondoftwo}%
\providecommand \bibfield [0]{\@secondoftwo}%
\providecommand \translation [1]{[#1]}%
\providecommand \BibitemOpen[0]{}%
\providecommand \bibitemStop [0]{}%
\providecommand \bibitemNoStop [0]{.\EOS\space}%
\providecommand \EOS [0]{\spacefactor3000\relax}%
\providecommand \BibitemShut [1]{\csname bibitem#1\endcsname}%
\bibitem{lammert1996ion}%
  \BibitemOpen
  \bibfield{author}{%
  \bibinfo {author} {\bibfnamefont{P.~E.}\ \bibnamefont{Lammert}}, \bibinfo
  {author} {\bibfnamefont{J.}~\bibnamefont{Prost}},\ and\ \bibinfo {author}
  {\bibfnamefont{R.}~\bibnamefont{Bruinsma}},\ }%
  \bibfield{journal}{%
  \bibinfo {journal} {J. Theor. Biol.}\ }%
  \textbf{\bibinfo {volume} {178}},\ \bibinfo {pages} {387} (\bibinfo {year}
  {1996})%
  \bibAnnoteFile{NoStop}{lammert1996ion}%
\bibitem{imagilov2001}%
  \BibitemOpen
  \bibfield{author}{%
  \bibinfo {author} {\bibfnamefont{R.}~\bibnamefont{Ismagilov}}, \bibinfo
  {author} {\bibfnamefont{A.}~\bibnamefont{Schwartz}}, \bibinfo {author}
  {\bibfnamefont{N.}~\bibnamefont{Bowden}},\ and\ \bibinfo {author}
  {\bibfnamefont{G.}~\bibnamefont{Whitesides}},\ }%
  \bibfield{journal}{%
  \bibinfo {journal} {{Angew. Chem. Int. Ed.}}\ }%
  \textbf{\bibinfo {volume} {{41}}},\ \bibinfo {pages} {{652}} (\bibinfo {year}
  {{2002}})%
  \bibAnnoteFile{NoStop}{imagilov2001}%
\bibitem{paxton2005motility}%
  \BibitemOpen
  \bibfield{author}{%
  \bibinfo {author} {\bibfnamefont{W.~F.}\ \bibnamefont{Paxton}}, \bibinfo
  {author} {\bibfnamefont{A.}~\bibnamefont{Sen}},\ and\ \bibinfo {author}
  {\bibfnamefont{T.~E.}\ \bibnamefont{Mallouk}},\ }%
  \bibfield{journal}{%
  \bibinfo {journal} {Chem. Eur. J.}\ }%
  \textbf{\bibinfo {volume} {11}} (\bibinfo {year} {2005})%
  \bibAnnoteFile{NoStop}{paxton2005motility}%
\bibitem{golestanian2005propulsion}%
  \BibitemOpen
  \bibfield{author}{%
  \bibinfo {author} {\bibfnamefont{R.}~\bibnamefont{Golestanian}}, \bibinfo
  {author} {\bibfnamefont{T.~B.}\ \bibnamefont{Liverpool}},\ and\ \bibinfo
  {author} {\bibfnamefont{A.}~\bibnamefont{Ajdari}},\ }%
  \bibfield{journal}{%
  \bibinfo {journal} {Phys. Rev. Lett.}\ }%
  \textbf{\bibinfo {volume} {94}},\ \bibinfo {pages} {220801} (\bibinfo {year}
  {2005})%
  \bibAnnoteFile{NoStop}{golestanian2005propulsion}%
\bibitem{wang2006bipolar}%
  \BibitemOpen
  \bibfield{author}{%
  \bibinfo {author} {\bibfnamefont{Y.}~\bibnamefont{Wang}}, \bibinfo {author}
  {\bibfnamefont{R.}~\bibnamefont{Hernandez}}, \bibinfo {author}
  {\bibfnamefont{D.}~\bibnamefont{Bartlett~Jr}}, \bibinfo {author}
  {\bibfnamefont{J.}~\bibnamefont{Bingham}}, \bibinfo {author}
  {\bibfnamefont{T.}~\bibnamefont{Kline}}, \bibinfo {author}
  {\bibfnamefont{A.}~\bibnamefont{Sen}},\ and\ \bibinfo {author}
  {\bibfnamefont{T.}~\bibnamefont{Mallouk}},\ }%
  \bibfield{journal}{%
  \bibinfo {journal} {Langmuir}\ }%
  \textbf{\bibinfo {volume} {22}},\ \bibinfo {pages} {10451} (\bibinfo {year}
  {2006})%
  \bibAnnoteFile{NoStop}{wang2006bipolar}%
\bibitem{rückner2007chemically}%
  \BibitemOpen
  \bibfield{author}{%
  \bibinfo {author} {\bibfnamefont{G.}~\bibnamefont{R{\"u}ckner}}\ and\
  \bibinfo {author} {\bibfnamefont{R.}~\bibnamefont{Kapral}},\ }%
  \bibfield{journal}{%
  \bibinfo {journal} {Phys.Rev. Lett.}\ }%
  \textbf{\bibinfo {volume} {98}},\ \bibinfo {pages} {150603} (\bibinfo {year}
  {2007})%
  \bibAnnoteFile{NoStop}{rückner2007chemically}%
\bibitem{thutupalli2011simple}%
  \BibitemOpen
  \bibfield{author}{%
  \bibinfo {author} {\bibfnamefont{S.}~\bibnamefont{Thutupalli}}, \bibinfo
  {author} {\bibfnamefont{R.}~\bibnamefont{Seemann}},\ and\ \bibinfo {author}
  {\bibfnamefont{S.}~\bibnamefont{Herminghaus}},\ }%
  \bibfield{journal}{%
  \bibinfo {journal} {New J. Phys.}\ }%
  \textbf{\bibinfo {volume} {13}},\ \bibinfo {pages} {073021} (\bibinfo {year}
  {2011})%
  \bibAnnoteFile{NoStop}{thutupalli2011simple}%
\bibitem{thakur2011dynamics}%
  \BibitemOpen
  \bibfield{author}{%
  \bibinfo {author} {\bibfnamefont{S.}~\bibnamefont{Thakur}}\ and\ \bibinfo
  {author} {\bibfnamefont{R.}~\bibnamefont{Kapral}},\ }%
  \bibfield{journal}{%
  \bibinfo {journal} {J. Chem. Phys.}\ }%
  \textbf{\bibinfo {volume} {135}},\ \bibinfo {pages} {024509} (\bibinfo {year}
  {2011})%
  \bibAnnoteFile{NoStop}{thakur2011dynamics}%
\bibitem{ozin2005dream}%
  \BibitemOpen
  \bibfield{author}{%
  \bibinfo {author} {\bibfnamefont{G.}~\bibnamefont{Ozin}}, \bibinfo {author}
  {\bibfnamefont{I.}~\bibnamefont{Manners}}, \bibinfo {author}
  {\bibfnamefont{S.}~\bibnamefont{Fournier-Bidoz}},\ and\ \bibinfo {author}
  {\bibfnamefont{A.}~\bibnamefont{Arsenault}},\ }%
  \bibfield{journal}{%
  \bibinfo {journal} {Adv. Mater.}\ }%
  \textbf{\bibinfo {volume} {17}},\ \bibinfo {pages} {3011} (\bibinfo {year}
  {2005})%
  \bibAnnoteFile{NoStop}{ozin2005dream}%
\bibitem{paxton2006chemical}%
  \BibitemOpen
  \bibfield{author}{%
  \bibinfo {author} {\bibfnamefont{W.}~\bibnamefont{Paxton}}, \bibinfo {author}
  {\bibfnamefont{S.}~\bibnamefont{Sundararajan}}, \bibinfo {author}
  {\bibfnamefont{T.}~\bibnamefont{Mallouk}},\ and\ \bibinfo {author}
  {\bibfnamefont{A.}~\bibnamefont{Sen}},\ }%
  \bibfield{journal}{%
  \bibinfo {journal} {Angew. Chem. Int. Ed.}\ }%
  \textbf{\bibinfo {volume} {45}},\ \bibinfo {pages} {5420} (\bibinfo {year}
  {2006})%
  \bibAnnoteFile{NoStop}{paxton2006chemical}%
\bibitem{He2007}%
  \BibitemOpen
  \bibfield{author}{%
  \bibinfo {author} {\bibfnamefont{Y.}~\bibnamefont{He}}, \bibinfo {author}
  {\bibfnamefont{J.}~\bibnamefont{Wu}},\ and\ \bibinfo {author}
  {\bibfnamefont{Y.}~\bibnamefont{Zhao}},\ }%
  \bibfield{journal}{%
  \Doi{10.1021/nl070461j}{\bibinfo {journal} {Nano Lett.}}\ }%
  \textbf{\bibinfo {volume} {7}},\ \bibinfo {pages} {1369} (\bibinfo {year}
  {2007})%
  \bibAnnoteFile{NoStop}{He2007}%
\bibitem{burdick2008synthetic}%
  \BibitemOpen
  \bibfield{author}{%
  \bibinfo {author} {\bibfnamefont{J.}~\bibnamefont{Burdick}}, \bibinfo
  {author} {\bibfnamefont{R.}~\bibnamefont{Laocharoensuk}}, \bibinfo {author}
  {\bibfnamefont{P.}~\bibnamefont{Wheat}}, \bibinfo {author}
  {\bibfnamefont{J.}~\bibnamefont{Posner}},\ and\ \bibinfo {author}
  {\bibfnamefont{J.}~\bibnamefont{Wang}},\ }%
  \bibfield{journal}{%
  \bibinfo {journal} {J. Am. Chem. Soc.}\ }%
  \textbf{\bibinfo {volume} {130}},\ \bibinfo {pages} {8164} (\bibinfo {year}
  {2008})%
  \bibAnnoteFile{NoStop}{burdick2008synthetic}%
\bibitem{kaganrapid}%
  \BibitemOpen
  \bibfield{author}{%
  \bibinfo {author} {\bibfnamefont{D.}~\bibnamefont{Kagan}}, \bibinfo {author}
  {\bibfnamefont{R.}~\bibnamefont{Laocharoensuk}}, \bibinfo {author}
  {\bibfnamefont{M.}~\bibnamefont{Zimmerman}}, \bibinfo {author}
  {\bibfnamefont{C.}~\bibnamefont{Clawson}}, \bibinfo {author}
  {\bibfnamefont{S.}~\bibnamefont{Balasubramanian}}, \bibinfo {author}
  {\bibfnamefont{D.}~\bibnamefont{Kagan}}, \bibinfo {author}
  {\bibfnamefont{D.}~\bibnamefont{Bishop}}, \bibinfo {author}
  {\bibfnamefont{S.}~\bibnamefont{Sattayasamitsathit}}, \bibinfo {author}
  {\bibfnamefont{L.}~\bibnamefont{Zhang}},\ and\ \bibinfo {author}
  {\bibfnamefont{J.}~\bibnamefont{Wang}},\ }%
  \bibfield{journal}{%
  \bibinfo {journal} {Small}\ }%
  \textbf{\bibinfo {volume} {6}},\ \bibinfo {pages} {2741} (\bibinfo {year}
  {2010})%
  \bibAnnoteFile{NoStop}{kaganrapid}%
\bibitem{solovev2010magnetic}%
  \BibitemOpen
  \bibfield{author}{%
  \bibinfo {author} {\bibfnamefont{A.}~\bibnamefont{Solovev}}, \bibinfo
  {author} {\bibfnamefont{S.}~\bibnamefont{Sanchez}}, \bibinfo {author}
  {\bibfnamefont{M.}~\bibnamefont{Pumera}}, \bibinfo {author}
  {\bibfnamefont{Y.}~\bibnamefont{Mei}},\ and\ \bibinfo {author}
  {\bibfnamefont{O.}~\bibnamefont{Schmidt}},\ }%
  \bibfield{journal}{%
  \bibinfo {journal} {Adv. Funct. Mater.}\ }%
  \textbf{\bibinfo {volume} {20}},\ \bibinfo {pages} {2430} (\bibinfo {year}
  {2010})%
  \bibAnnoteFile{NoStop}{solovev2010magnetic}%
\bibitem{popescu2011pulling}%
  \BibitemOpen
  \bibfield{author}{%
  \bibinfo {author} {\bibfnamefont{M.}~\bibnamefont{Popescu}}, \bibinfo
  {author} {\bibfnamefont{M.}~\bibnamefont{Tasinkevych}},\ and\ \bibinfo
  {author} {\bibfnamefont{S.}~\bibnamefont{Dietrich}},\ }%
  \bibfield{journal}{%
  \bibinfo {journal} {Europhys. Lett.}\ }%
  \textbf{\bibinfo {volume} {95}},\ \bibinfo {pages} {28004} (\bibinfo {year}
  {2011})%
  \bibAnnoteFile{NoStop}{popescu2011pulling}%
\bibitem{golestanian2009anomalous}%
  \BibitemOpen
  \bibfield{author}{%
  \bibinfo {author} {\bibfnamefont{R.}~\bibnamefont{Golestanian}},\ }%
  \bibfield{journal}{%
  \bibinfo {journal} {Phys. Rev. Lett.}\ }%
  \textbf{\bibinfo {volume} {102}},\ \bibinfo {pages} {188305} (\bibinfo {year}
  {2009})%
  \bibAnnoteFile{NoStop}{golestanian2009anomalous}%
\bibitem{campos2009superdiffusive}%
  \BibitemOpen
  \bibfield{author}{%
  \bibinfo {author} {\bibfnamefont{D.}~\bibnamefont{Campos}}\ and\ \bibinfo
  {author} {\bibfnamefont{V.}~\bibnamefont{M{\'e}ndez}},\ }%
  \bibfield{journal}{%
  \bibinfo {journal} {J. Chem. Phys.}\ }%
  \textbf{\bibinfo {volume} {130}},\ \bibinfo {pages} {134711} (\bibinfo {year}
  {2009})%
  \bibAnnoteFile{NoStop}{campos2009superdiffusive}%
\bibitem{ten2011brownian}%
  \BibitemOpen
  \bibfield{author}{%
  \bibinfo {author} {\bibfnamefont{B.}~\bibnamefont{ten Hagen}}, \bibinfo
  {author} {\bibfnamefont{S.}~\bibnamefont{van Teeffelen}},\ and\ \bibinfo
  {author} {\bibfnamefont{H.}~\bibnamefont{L\"owen}},\ }%
  \bibfield{journal}{%
  \bibinfo {journal} {J. Phys.: Condens. Matter}\ }%
  \textbf{\bibinfo {volume} {23}},\ \bibinfo {pages} {194119} (\bibinfo {year}
  {2011})%
  \bibAnnoteFile{NoStop}{ten2011brownian}%
\bibitem{palacci2010sedimentation}%
  \BibitemOpen
  \bibfield{author}{%
  \bibinfo {author} {\bibfnamefont{J.}~\bibnamefont{Palacci}}, \bibinfo
  {author} {\bibfnamefont{C.}~\bibnamefont{Cottin-Bizonne}}, \bibinfo {author}
  {\bibfnamefont{C.}~\bibnamefont{Ybert}},\ and\ \bibinfo {author}
  {\bibfnamefont{L.}~\bibnamefont{Bocquet}},\ }%
  \bibfield{journal}{%
  \bibinfo {journal} {Phys. Rev. Lett.}\ }%
  \textbf{\bibinfo {volume} {105}},\ \bibinfo {pages} {88304} (\bibinfo {year}
  {2010})%
  \bibAnnoteFile{NoStop}{palacci2010sedimentation}%
\bibitem{sano2011}%
  \BibitemOpen
  \bibfield{author}{%
  \bibinfo {author} {\bibfnamefont{H.~R.}\ \bibnamefont{Suzuki},
  \bibfnamefont{R.~Jiang}}\ and\ \bibinfo {author}
  {\bibfnamefont{M.}~\bibnamefont{Sano}},\ }%
  \bibfield{journal}{%
  \bibinfo {journal} {Arxiv preprint arXiv:1104.5607v1}}%
   (\bibinfo {year} {2011})%
  \bibAnnoteFile{NoStop}{sano2011}%
\bibitem{popescu2009confinement}%
  \BibitemOpen
  \bibfield{author}{%
  \bibinfo {author} {\bibfnamefont{M.}~\bibnamefont{Popescu}}, \bibinfo
  {author} {\bibfnamefont{S.}~\bibnamefont{Dietrich}},\ and\ \bibinfo {author}
  {\bibfnamefont{G.}~\bibnamefont{Oshanin}},\ }%
  \bibfield{journal}{%
  \bibinfo {journal} {J. Chem. Phys.}\ }%
  \textbf{\bibinfo {volume} {130}},\ \bibinfo {pages} {194702} (\bibinfo {year}
  {2009})%
  \bibAnnoteFile{NoStop}{popescu2009confinement}%
\bibitem{yang2010swarm}%
  \BibitemOpen
  \bibfield{author}{%
  \bibinfo {author} {\bibfnamefont{Y.}~\bibnamefont{Yang}}, \bibinfo {author}
  {\bibfnamefont{V.}~\bibnamefont{Marceau}},\ and\ \bibinfo {author}
  {\bibfnamefont{G.}~\bibnamefont{Gompper}},\ }%
  \bibfield{journal}{%
  \bibinfo {journal} {Phys. Rev. E}\ }%
  \textbf{\bibinfo {volume} {82}},\ \bibinfo {pages} {031904} (\bibinfo {year}
  {2010})%
  \bibAnnoteFile{NoStop}{yang2010swarm}%
\bibitem{enculescu2011}%
  \BibitemOpen
  \bibfield{author}{%
  \bibinfo {author} {\bibfnamefont{M.}~\bibnamefont{Enculescu}}\ and\ \bibinfo
  {author} {\bibfnamefont{H.}~\bibnamefont{Stark}},\ }%
  \bibfield{journal}{%
  \Doi{10.1103/PhysRevLett.107.058301}{\bibinfo {journal} {Phys. Rev. Lett.}}\
  }%
  \textbf{\bibinfo {volume} {107}},\ \bibinfo {pages} {058301} (\bibinfo {year}
  {2011})%
  \bibAnnoteFile{NoStop}{enculescu2011}%
\bibitem{thakurinteraction}%
  \BibitemOpen
  \bibfield{author}{%
  \bibinfo {author} {\bibfnamefont{S.}~\bibnamefont{Thakur}}, \bibinfo {author}
  {\bibfnamefont{J.}~\bibnamefont{Chen}},\ and\ \bibinfo {author}
  {\bibfnamefont{R.}~\bibnamefont{Kapral}},\ }%
  \bibfield{journal}{%
  \bibinfo {journal} {Angew. Chem. Int. Ed.}\ }%
  \textbf{\bibinfo {volume} {50}},\ \bibinfo {pages} {1} (\bibinfo {year}
  {2011})%
  \bibAnnoteFile{NoStop}{thakurinteraction}%
\bibitem{hall1997electrochemical}%
  \BibitemOpen
  \bibfield{author}{%
  \bibinfo {author} {\bibfnamefont{S.}~\bibnamefont{Hall}}, \bibinfo {author}
  {\bibfnamefont{E.}~\bibnamefont{Khudaish}},\ and\ \bibinfo {author}
  {\bibfnamefont{A.}~\bibnamefont{Hart}},\ }%
  \bibfield{journal}{%
  \bibinfo {journal} {Electrochim. acta}\ }%
  \textbf{\bibinfo {volume} {43}},\ \bibinfo {pages} {579} (\bibinfo {year}
  {1997})%
  \bibAnnoteFile{NoStop}{hall1997electrochemical}%
\bibitem{paxton2006catalytically}%
  \BibitemOpen
  \bibfield{author}{%
  \bibinfo {author} {\bibfnamefont{W.~F.}\ \bibnamefont{Paxton}}, \bibinfo
  {author} {\bibfnamefont{P.~T.}\ \bibnamefont{Baker}}, \bibinfo {author}
  {\bibfnamefont{T.~R.}\ \bibnamefont{Kline}}, \bibinfo {author}
  {\bibfnamefont{Y.}~\bibnamefont{Wang}}, \bibinfo {author}
  {\bibfnamefont{T.~E.}\ \bibnamefont{Mallouk}},\ and\ \bibinfo {author}
  {\bibfnamefont{A.}~\bibnamefont{Sen}},\ }%
  \bibfield{journal}{%
  \bibinfo {journal} {J. Am. Chem. Soc.}\ }%
  \textbf{\bibinfo {volume} {128}},\ \bibinfo {pages} {14881} (\bibinfo {year}
  {2006})%
  \bibAnnoteFile{NoStop}{paxton2006catalytically}%
\bibitem{sundararajan2008catalytic}%
  \BibitemOpen
  \bibfield{author}{%
  \bibinfo {author} {\bibfnamefont{S.}~\bibnamefont{Sundararajan}}, \bibinfo
  {author} {\bibfnamefont{P.}~\bibnamefont{Lammert}}, \bibinfo {author}
  {\bibfnamefont{A.}~\bibnamefont{Zudans}}, \bibinfo {author}
  {\bibfnamefont{V.}~\bibnamefont{Crespi}},\ and\ \bibinfo {author}
  {\bibfnamefont{A.}~\bibnamefont{Sen}},\ }%
  \bibfield{journal}{%
  \bibinfo {journal} {Nano Lett.}\ }%
  \textbf{\bibinfo {volume} {8}},\ \bibinfo {pages} {1271} (\bibinfo {year}
  {2008})%
  \bibAnnoteFile{NoStop}{sundararajan2008catalytic}%
\bibitem{moran2010locomotion}%
  \BibitemOpen
  \bibfield{author}{%
  \bibinfo {author} {\bibfnamefont{J.}~\bibnamefont{Moran}}, \bibinfo {author}
  {\bibfnamefont{P.}~\bibnamefont{Wheat}},\ and\ \bibinfo {author}
  {\bibfnamefont{J.}~\bibnamefont{Posner}},\ }%
  \bibfield{journal}{%
  \bibinfo {journal} {Phys. Rev. E}\ }%
  \textbf{\bibinfo {volume} {81}},\ \bibinfo {pages} {065302} (\bibinfo {year}
  {2010})%
  \bibAnnoteFile{NoStop}{moran2010locomotion}%
\bibitem{yariv2010electrokinetic}%
  \BibitemOpen
  \bibfield{author}{%
  \bibinfo {author} {\bibfnamefont{E.}~\bibnamefont{Yariv}},\ }%
  \bibfield{journal}{%
  \bibinfo {journal} {Proc. R. Soc. A}\ }%
  \textbf{\bibinfo {volume} {467}},\ \bibinfo {pages} {1645} (\bibinfo {year}
  {2011})%
  \bibAnnoteFile{NoStop}{yariv2010electrokinetic}%
\bibitem{anderson1989colloid}%
  \BibitemOpen
  \bibfield{author}{%
  \bibinfo {author} {\bibfnamefont{J.~L.}\ \bibnamefont{Anderson}},\ }%
  \bibfield{journal}{%
  \bibinfo {journal} {Annu. Rev. Fluid Mech.}\ }%
  \textbf{\bibinfo {volume} {21}},\ \bibinfo {pages} {61} (\bibinfo {year}
  {1989})%
  \bibAnnoteFile{NoStop}{anderson1989colloid}%
\bibitem{paxton2004}%
  \BibitemOpen
  \bibfield{author}{%
  \bibinfo {author} {\bibfnamefont{W.~F.}\ \bibnamefont{Paxton}}, \bibinfo
  {author} {\bibfnamefont{K.~C.}\ \bibnamefont{Kistler}}, \bibinfo {author}
  {\bibfnamefont{C.~C.}\ \bibnamefont{Olmeda}}, \bibinfo {author}
  {\bibfnamefont{A.}~\bibnamefont{Sen}}, \bibinfo {author}
  {\bibfnamefont{S.~K.}\ \bibnamefont{St.~Angelo}}, \bibinfo {author}
  {\bibfnamefont{Y.}~\bibnamefont{Cao}}, \bibinfo {author}
  {\bibfnamefont{T.~E.}\ \bibnamefont{Mallouk}}, \bibinfo {author}
  {\bibfnamefont{P.~E.}\ \bibnamefont{Lammert}},\ and\ \bibinfo {author}
  {\bibfnamefont{V.~H.}\ \bibnamefont{Crespi}},\ }%
  \bibfield{journal}{%
  \Doi{10.1021/ja047697z}{\bibinfo {journal} {J. Am. Chem. Soc.}}\ }%
  \textbf{\bibinfo {volume} {126}},\ \bibinfo {pages} {13424} (\bibinfo {year}
  {2004})%
  \bibAnnoteFile{NoStop}{paxton2004}%
\bibitem{gibbs2009autonomously}%
  \BibitemOpen
  \bibfield{author}{%
  \bibinfo {author} {\bibfnamefont{J.}~\bibnamefont{Gibbs}}\ and\ \bibinfo
  {author} {\bibfnamefont{Y.}~\bibnamefont{Zhao}},\ }%
  \bibfield{journal}{%
  \bibinfo {journal} {Appl. Phys. Lett.}\ }%
  \textbf{\bibinfo {volume} {94}},\ \bibinfo {pages} {163104} (\bibinfo {year}
  {2009})%
  \bibAnnoteFile{NoStop}{gibbs2009autonomously}%
\bibitem{howse2007self}%
  \BibitemOpen
  \bibfield{author}{%
  \bibinfo {author} {\bibfnamefont{J.~R.}\ \bibnamefont{Howse}}, \bibinfo
  {author} {\bibfnamefont{R.~A.~L.}\ \bibnamefont{Jones}}, \bibinfo {author}
  {\bibfnamefont{A.~J.}\ \bibnamefont{Ryan}}, \bibinfo {author}
  {\bibfnamefont{T.}~\bibnamefont{Gough}}, \bibinfo {author}
  {\bibfnamefont{R.}~\bibnamefont{Vafabakhsh}},\ and\ \bibinfo {author}
  {\bibfnamefont{R.}~\bibnamefont{Golestanian}},\ }%
  \bibfield{journal}{%
  \bibinfo {journal} {Phys. Rev. Lett.}\ }%
  \textbf{\bibinfo {volume} {99}},\ \bibinfo {pages} {48102} (\bibinfo {year}
  {2007})%
  \bibAnnoteFile{NoStop}{howse2007self}%
\bibitem{shi2009computational}%
  \BibitemOpen
  \bibfield{author}{%
  \bibinfo {author} {\bibfnamefont{Y.}~\bibnamefont{Shi}}, \bibinfo {author}
  {\bibfnamefont{L.}~\bibnamefont{Huang}},\ and\ \bibinfo {author}
  {\bibfnamefont{D.}~\bibnamefont{Brenner}},\ }%
  \bibfield{journal}{%
  \bibinfo {journal} {J. Chem. Phys.}\ }%
  \textbf{\bibinfo {volume} {131}},\ \bibinfo {pages} {014705} (\bibinfo {year}
  {2009})%
  \bibAnnoteFile{NoStop}{shi2009computational}%
\bibitem{sabass2010efficiency}%
  \BibitemOpen
  \bibfield{author}{%
  \bibinfo {author} {\bibfnamefont{B.}~\bibnamefont{Sabass}}\ and\ \bibinfo
  {author} {\bibfnamefont{U.}~\bibnamefont{Seifert}},\ }%
  \bibfield{journal}{%
  \bibinfo {journal} {Phys. Rev. Lett.}\ }%
  \textbf{\bibinfo {volume} {105}},\ \bibinfo {pages} {218103} (\bibinfo {year}
  {2010})%
  \bibAnnoteFile{NoStop}{sabass2010efficiency}%
\bibitem{spagnolie2010optimal}%
  \BibitemOpen
  \bibfield{author}{%
  \bibinfo {author} {\bibfnamefont{S.}~\bibnamefont{Spagnolie}}\ and\ \bibinfo
  {author} {\bibfnamefont{E.}~\bibnamefont{Lauga}},\ }%
  \bibfield{journal}{%
  \bibinfo {journal} {Phys. Fluids}\ }%
  \textbf{\bibinfo {volume} {22}},\ \bibinfo {pages} {031901} (\bibinfo {year}
  {2010})%
  \bibAnnoteFile{NoStop}{spagnolie2010optimal}%
\bibitem{teubner1982motion}%
  \BibitemOpen
  \bibfield{author}{%
  \bibinfo {author} {\bibfnamefont{M.}~\bibnamefont{Teubner}},\ }%
  \bibfield{journal}{%
  \bibinfo {journal} {J. Chem. Phys.}\ }%
  \textbf{\bibinfo {volume} {76}},\ \bibinfo {pages} {5564} (\bibinfo {year}
  {1982})%
  \bibAnnoteFile{NoStop}{teubner1982motion}%
\bibitem{o1978electrophoretic}%
  \BibitemOpen
  \bibfield{author}{%
  \bibinfo {author} {\bibfnamefont{R.}~\bibnamefont{O'Brien}}\ and\ \bibinfo
  {author} {\bibfnamefont{L.}~\bibnamefont{White}},\ }%
  \bibfield{journal}{%
  \bibinfo {journal} {J. Chem. Soc. Faraday Trans. 2}\ }%
  \textbf{\bibinfo {volume} {74}},\ \bibinfo {pages} {1607} (\bibinfo {year}
  {1978})%
  \bibAnnoteFile{NoStop}{o1978electrophoretic}%
\bibitem{happel1983martinus}%
  \BibitemOpen
  \bibfield{author}{%
  \bibinfo {author} {\bibfnamefont{J.}~\bibnamefont{Happel}}\ and\ \bibinfo
  {author} {\bibfnamefont{H.}~\bibnamefont{Brenner}},\ }%
  \emph{\bibinfo {title} {{Low Reynolds number hydrodynamics}}}\ (\bibinfo
  {publisher} {Martinus Nijhoff},\ \bibinfo {year} {1983})%
  \bibAnnoteFile{NoStop}{happel1983martinus}%
\bibitem{acrivos1962heat}%
  \BibitemOpen
  \bibfield{author}{%
  \bibinfo {author} {\bibfnamefont{A.}~\bibnamefont{Acrivos}}\ and\ \bibinfo
  {author} {\bibfnamefont{T.~D.}\ \bibnamefont{Taylor}},\ }%
  \bibfield{journal}{%
  \bibinfo {journal} {Phys. Fluids}\ }%
  \textbf{\bibinfo {volume} {5}},\ \bibinfo {pages} {387} (\bibinfo {year}
  {1962})%
  \bibAnnoteFile{NoStop}{acrivos1962heat}%
\bibitem{keh2000diffusiophoretic}%
  \BibitemOpen
  \bibfield{author}{%
  \bibinfo {author} {\bibfnamefont{H.~J.}\ \bibnamefont{Keh}}\ and\ \bibinfo
  {author} {\bibfnamefont{Y.~K.}\ \bibnamefont{Wei}},\ }%
  \bibfield{journal}{%
  \bibinfo {journal} {Langmuir}\ }%
  \textbf{\bibinfo {volume} {16}},\ \bibinfo {pages} {5289} (\bibinfo {year}
  {2000})%
  \bibAnnoteFile{NoStop}{keh2000diffusiophoretic}%
\bibitem{anderson1982motion}%
  \BibitemOpen
  \bibfield{author}{%
  \bibinfo {author} {\bibfnamefont{J.~L.}\ \bibnamefont{Anderson}}, \bibinfo
  {author} {\bibfnamefont{M.~E.}\ \bibnamefont{Lowell}},\ and\ \bibinfo
  {author} {\bibfnamefont{D.~C.}\ \bibnamefont{Prieve}},\ }%
  \bibfield{journal}{%
  \bibinfo {journal} {J. Fluid Mech.}\ }%
  \textbf{\bibinfo {volume} {117}} (\bibinfo {year} {1982})%
  \bibAnnoteFile{NoStop}{anderson1982motion}%
\bibitem{saville1977electrokinetic}%
  \BibitemOpen
  \bibfield{author}{%
  \bibinfo {author} {\bibfnamefont{D.}~\bibnamefont{Saville}},\ }%
  \bibfield{journal}{%
  \bibinfo {journal} {Annu. Rev. Fluid Mech.}\ }%
  \textbf{\bibinfo {volume} {9}},\ \bibinfo {pages} {321} (\bibinfo {year}
  {1977})%
  \bibAnnoteFile{NoStop}{saville1977electrokinetic}%
\bibitem{prieve1987diffusiophoresis}%
  \BibitemOpen
  \bibfield{author}{%
  \bibinfo {author} {\bibfnamefont{D.}~\bibnamefont{Prieve}}\ and\ \bibinfo
  {author} {\bibfnamefont{R.}~\bibnamefont{Roman}},\ }%
  \bibfield{journal}{%
  \bibinfo {journal} {J. Chem. Soc., Faraday Trans. 2}\ }%
  \textbf{\bibinfo {volume} {83}},\ \bibinfo {pages} {1287} (\bibinfo {year}
  {1987})%
  \bibAnnoteFile{NoStop}{prieve1987diffusiophoresis}%
\bibitem{anderson1991diffusiophoresis}%
  \BibitemOpen
  \bibfield{author}{%
  \bibinfo {author} {\bibfnamefont{J.~L.}\ \bibnamefont{Anderson}}\ and\
  \bibinfo {author} {\bibfnamefont{D.~C.}\ \bibnamefont{Prieve}},\ }%
  \bibfield{journal}{%
  \bibinfo {journal} {Langmuir}\ }%
  \textbf{\bibinfo {volume} {7}},\ \bibinfo {pages} {403} (\bibinfo {year}
  {1991})%
  \bibAnnoteFile{NoStop}{anderson1991diffusiophoresis}%
\bibitem{park2008concentration}%
  \BibitemOpen
  \bibfield{author}{%
  \bibinfo {author} {\bibfnamefont{S.}~\bibnamefont{Park}}\ and\ \bibinfo
  {author} {\bibfnamefont{N.}~\bibnamefont{Agmon}},\ }%
  \bibfield{journal}{%
  \bibinfo {journal} {J. Phys. Chem. B}\ }%
  \textbf{\bibinfo {volume} {112}},\ \bibinfo {pages} {12104} (\bibinfo {year}
  {2008})%
  \bibAnnoteFile{NoStop}{park2008concentration}%
\bibitem{solc1973kinetics}%
  \BibitemOpen
  \bibfield{author}{%
  \bibinfo {author} {\bibfnamefont{K.}~\bibnamefont{\v{S}olc}}\ and\ \bibinfo
  {author} {\bibfnamefont{W.}~\bibnamefont{Stockmayer}},\ }%
  \bibfield{journal}{%
  \Doi{10.1063/1.1675283}{\bibinfo {journal} {J. Chem. Phys.}}\ }%
  \textbf{\bibinfo {volume} {54}},\ \bibinfo {pages} {2981} (\bibinfo {year}
  {1971})%
  \bibAnnoteFile{NoStop}{solc1973kinetics}%
\bibitem{shoup1981diffusion}%
  \BibitemOpen
  \bibfield{author}{%
  \bibinfo {author} {\bibfnamefont{D.}~\bibnamefont{Shoup}}, \bibinfo {author}
  {\bibfnamefont{G.}~\bibnamefont{Lipari}},\ and\ \bibinfo {author}
  {\bibfnamefont{A.}~\bibnamefont{Szabo}},\ }%
  \bibfield{journal}{%
  \bibinfo {journal} {Biophys. J.}\ }%
  \textbf{\bibinfo {volume} {36}},\ \bibinfo {pages} {697} (\bibinfo {year}
  {1981})%
  \bibAnnoteFile{NoStop}{shoup1981diffusion}%
\bibitem{lighthill1952squirming}%
  \BibitemOpen
  \bibfield{author}{%
  \bibinfo {author} {\bibfnamefont{M.~J.}\ \bibnamefont{Lighthill}},\ }%
  \bibfield{journal}{%
  \bibinfo {journal} {Commun. Pure Appl. Math.}\ }%
  \textbf{\bibinfo {volume} {5}} (\bibinfo {year} {1952})%
  \bibAnnoteFile{NoStop}{lighthill1952squirming}%
\bibitem{ke2010motion}%
  \BibitemOpen
  \bibfield{author}{%
  \bibinfo {author} {\bibfnamefont{H.}~\bibnamefont{Ke}}, \bibinfo {author}
  {\bibfnamefont{S.}~\bibnamefont{Ye}}, \bibinfo {author}
  {\bibfnamefont{R.}~\bibnamefont{Carroll}},\ and\ \bibinfo {author}
  {\bibfnamefont{K.}~\bibnamefont{Showalter}},\ }%
  \bibfield{journal}{%
  \bibinfo {journal} {J. Phys. Chem. A}\ }%
  \textbf{\bibinfo {volume} {114}},\ \bibinfo {pages} {5462} (\bibinfo {year}
  {2010})%
  \bibAnnoteFile{NoStop}{ke2010motion}%
\bibitem{anderson1984diffusiophoresis}%
  \BibitemOpen
  \bibfield{author}{%
  \bibinfo {author} {\bibfnamefont{J.~L.}\ \bibnamefont{Anderson}}\ and\
  \bibinfo {author} {\bibfnamefont{D.~C.}\ \bibnamefont{Prieve}},\ }%
  \bibfield{journal}{%
  \bibinfo {journal} {Sep. Purif. Methods}\ }%
  \textbf{\bibinfo {volume} {13}},\ \bibinfo {pages} {67} (\bibinfo {year}
  {1984})%
  \bibAnnoteFile{NoStop}{anderson1984diffusiophoresis}%
\bibitem{parmeggiani1999energy}%
  \BibitemOpen
  \bibfield{author}{%
  \bibinfo {author} {\bibfnamefont{A.}~\bibnamefont{Parmeggiani}}, \bibinfo
  {author} {\bibfnamefont{F.}~\bibnamefont{J{\"u}licher}}, \bibinfo {author}
  {\bibfnamefont{A.}~\bibnamefont{Ajdari}},\ and\ \bibinfo {author}
  {\bibfnamefont{J.}~\bibnamefont{Prost}},\ }%
  \bibfield{journal}{%
  \bibinfo {journal} {Phys. Rev. E}\ }%
  \textbf{\bibinfo {volume} {60}},\ \bibinfo {pages} {2127} (\bibinfo {year}
  {1999})%
  \bibAnnoteFile{NoStop}{parmeggiani1999energy}%
\bibitem{ajdari2006giant}%
  \BibitemOpen
  \bibfield{author}{%
  \bibinfo {author} {\bibfnamefont{A.}~\bibnamefont{Ajdari}}\ and\ \bibinfo
  {author} {\bibfnamefont{L.}~\bibnamefont{Bocquet}},\ }%
  \bibfield{journal}{%
  \bibinfo {journal} {Phys. Rev. Lett.}\ }%
  \textbf{\bibinfo {volume} {96}},\ \bibinfo {pages} {186102} (\bibinfo {year}
  {2006})%
  \bibAnnoteFile{NoStop}{ajdari2006giant}%
\bibitem{van1975perturbation}%
  \BibitemOpen
  \bibfield{author}{%
  \bibinfo {author} {\bibfnamefont{M.}~\bibnamefont{Van~Dyke}},\ }%
  \emph{\bibinfo {title} {Perturbation methods in fluid mechanics}}\ (\bibinfo
  {publisher} {Parabolic Press},\ \bibinfo {year} {1975})%
  \bibAnnoteFile{NoStop}{van1975perturbation}%
\bibitem{degroot1984}%
  \BibitemOpen
  \bibfield{author}{%
  \bibinfo {author} {\bibfnamefont{S.}~\bibnamefont{De~Groot}}\ and\ \bibinfo
  {author} {\bibfnamefont{P.}~\bibnamefont{Mazur}},\ }%
  \emph{\bibinfo {title} {{Non-equilibrium thermodynamics}}}\ (\bibinfo
  {publisher} {Dover publications},\ \bibinfo {year} {1984})%
  \bibAnnoteFile{NoStop}{degroot1984}%
\bibitem{julicher2009generic}%
  \BibitemOpen
  \bibfield{author}{%
  \bibinfo {author} {\bibfnamefont{F.}~\bibnamefont{J\"ulicher}}\ and\ \bibinfo
  {author} {\bibfnamefont{J.}~\bibnamefont{Prost}},\ }%
  \bibfield{journal}{%
  \bibinfo {journal} {Eur. Phys. J. E}\ }%
  \textbf{\bibinfo {volume} {29}},\ \bibinfo {pages} {27} (\bibinfo {year}
  {2009})%
  \bibAnnoteFile{NoStop}{julicher2009generic}%
\end{thebibliography}
%
\end{document}